%% file: neurips_2026.tex
\documentclass{article}

    \PassOptionsToPackage{numbers, compress}{natbib}

\usepackage[eandd, final]{neurips_2026}

\usepackage[utf8]{inputenc} %
\usepackage[T1]{fontenc}    %
\usepackage{hyperref}       %
\usepackage{url}            %
\usepackage{booktabs}       %
\usepackage{amsfonts}       %
\usepackage{nicefrac}       %
\usepackage{microtype}      %
\usepackage{xcolor}         %
\usepackage{soul}
\usepackage{tabularx}
\usepackage{xspace} 
\usepackage{subcaption}
\usepackage[figuresleft]{rotating}
\usepackage{multirow}
\usepackage[most]{tcolorbox}
\usepackage{todonotes}
\usepackage{inconsolata}
\usepackage{siunitx}
\usepackage{cleveref}

\newcommand{\consolasfont}{\ttfamily}
\newcommand{\smartparagraph}[1]{\vspace{.05in}\noindent\textbf{#1}}
    \newcommand{\namebenchmark}{LibEvoBench}

\newtcolorbox{promptbox}[1]{
    enhanced,
    coltitle=white,
    boxrule=0.5pt,
    colframe=black!75,            %
    colbacktitle=black!75,        %
    coltitle=white,            %
    colback=gray!10!white,     %
    title={#1}                 %
}

\title{\namebenchmark{}: Probing Temporal Knowledge Stratification in Code Generation Models}

\author{%
  \textbf{Daniele Cipollone\textsuperscript{1,2} \quad
  Sergey Titov\textsuperscript{2} \quad
  Maliheh Izadi\textsuperscript{1}} \\
  \textbf{Egor Bogomolov\textsuperscript{2} \quad
  Arie van Deursen\textsuperscript{1}}
  \\[0.6em]
  \normalfont
  \textsuperscript{1}Faculty of EEMCS, Delft University of Technology, Delft, Netherlands \\
  \textsuperscript{2}JetBrains Research, Amsterdam, Netherlands
}

\begin{document}

\maketitle

\begin{abstract}

Large software projects often depend on older versions of libraries, even as APIs continue to evolve across releases. This creates a challenge for LLMs: they must maintain knowledge of multiple API versions, not merely the latest or most common one. However, current LLMs are trained on temporally mixed corpora and lack explicit mechanisms for such version-specific reasoning, leading to anachronistic errors -- calling APIs as they exist in a different library \emph{version}. To systematically evaluate this phenomenon, we introduce \namebenchmark{}, a multi-task benchmark spanning multiple versions of widely used Python libraries, along with a new metric, the Software Evolution Understanding Score (SEUS), to measure models' consistency when working with evolving APIs. Our results show that state-of-the-art models are largely version-oblivious: performance degrades for evolving APIs, while for stable APIs it remains the same across versions. Moreover, simply specifying the target version provides no benefit, while relevant documentation significantly boosts models' accuracy. These findings highlight a systematic limitation of current training paradigms and motivate new approaches for temporally grounded knowledge in code generation.

\end{abstract}

\input{sections/001-intro}

\input{sections/111-method}

\input{sections/122-experiments}

\input{sections/222-related}

\input{sections/444-conclusion}

\begin{ack}
This work was conducted as part of the AI for Software Engineering (AI4SE) collaboration between JetBrains and Delft University of Technology, funded by JetBrains. The authors declare no competing interests beyond the affiliations listed.
\end{ack}
\bibliographystyle{plainnat}
\bibliography{references_main}

\newpage
\appendix
\input{sections/999-appendix}

\newpage
\input{checklist.tex}

\end{document}

%% file: sections/001-intro.tex
\section{Introduction}

Modern software development depends heavily on third-party libraries.
Production codebases often pin hundreds of independently versioned packages, each evolving on its own release cadence with new entry points, modified signatures, and deprecated symbols. As a result, version upgrades in large projects tend to be deliberate and infrequent, constrained by compatibility requirements and the cost of migrating downstream consumers~\cite{kula_developers_2018}.
As we show in Appendix~\ref{app:version-selection}, the library version used during development may lag significantly behind the latest available release. This gap creates a \emph{version-specific development context}: the correct API call, its expected parameters, and even its availability depend not on the library in general, but on the exact version specified in the project's dependency manifest.

While large language models (LLMs) have achieved remarkable results in programming tasks~\cite{lozhkov_starcoder_2024, team_gemini_2025, noauthor_introducing_nodate}, their training procedures do not account for version specificity. These models are trained on corpora that aggregate code and documentation from many library versions simultaneously. The training objective compresses this heterogeneous material into a single set of weights, with no explicit mechanism to stratify knowledge temporally.
Prior work has shown that language models struggle with temporally evolving factual knowledge~\cite{dhingra_time-aware_2022, wei_time_2025, zhu_evolvebench_2025}, and temporal alignment methods aim to condition retrieval on a target timestamp~\cite{zhao_set_2024}.
Version-aware code generation poses an even greater challenge: models must distinguish among multiple coexisting versions of the same library API, or risk producing \emph{anachronistic} code that is valid for one version but incorrect for the requested version.

To investigate how LLMs handle evolving library APIs, we introduce \textbf{\namebenchmark{}}, a multi-task benchmark spanning major versions of widely used Python libraries. 
\namebenchmark{} evaluates three tasks: API Calling (API-C), which tests contextual API use; API Identification (API-I), which tests whether models can identify the appropriate API from a description; and Signature Recall (SR), which tests version-specific signature knowledge.

To pinpoint the effects of API evolution on LLM behavior, we track public APIs and signatures across versions of each library in \namebenchmark{} to define two API subsets: \emph{stable} APIs, which remain unchanged across selected versions, and \emph{evolving} APIs, which are added, removed, or modified.
We then introduce the \emph{Software Evolution Understanding Score (SEUS)}, which measures a model’s ability to perform consistently across both stable and evolving APIs, penalizing inconsistencies along the version axis, and capturing anachronistic errors. This metric enables ranking models based on their robustness under varying library versions in software engineering scenarios.

We evaluate frontier proprietary models from OpenAI~\cite{noauthor_introducing_2026}, Anthropic~\cite{noauthor_introducing_nodate}, and Google~\cite{team_gemini_2025}, as well as the Qwen3.5 model family~\cite{team_qwen35-omni_2026}.
Across most models, accuracy on evolving APIs drops for newer library versions while stable-API performance stays nearly flat. Even the strongest models show a persistent 7--10 point accuracy gap between operating with stable and evolving APIs.
Within the Qwen3.5 family, models of varying sizes follow similar downward trends on the evolving curve. 
These results suggest that the degradation reflects a structural limitation of the training process rather than a limitation of model capability or scale.

We also find that most models are effectively version-oblivious. Providing the correct API documentation in the prompt yields a consistent 10–20 point improvement in API-calling accuracy, indicating that models can retrieve and use relevant information when it is made explicit. However, adding an explicit library version constraint produces no measurable improvement across any model family.
Our analysis suggests that the errors mostly stem from confusion rather than fabrication: models most often produce valid but incorrect symbols from the same library. At the parameter level, a persistent portion of these errors is \emph{anachronistic}: calls that would be correct for a different version of the same API.

\smartparagraph{Why parametric knowledge matters.}
We deliberately evaluate models in a single-step setting without tools, because this is the knowledge code generation actually runs on. Even inside an agent, a model emits hundreds of API calls per file in a single pass. It retrieves documentation or inspects installed packages only when it recognizes a gap in its knowledge. The failure we measure produces no such signal: a model unaware that an API has changed falls back on familiar APIs that still resolve and execute at the target version. The resulting code passes static checks and gives retrieval nothing to act on.

With this work, we make the following contributions. 
(i) We introduce \namebenchmark{}, a novel benchmark for evaluating how well LLMs can navigate evolving APIs of software libraries. 
(ii) We introduce \emph{SEUS}, a unified metric to analyze the robustness of LLMs to changes in software libraries. 
(iii) We use these tools to evaluate frontier and open-weight LLMs, showing that current models remain largely version-oblivious even when given explicit version constraints.
Together, these contributions provide a testbed for future researchers to develop temporally grounded and continual-learning methods for the evolving software engineering ecosystem.

%% file: sections/111-method.tex
\section{\namebenchmark{}}
\label{sec:benchmark}
\begin{figure*}[t]
  \centering
  \includegraphics[width=1\linewidth]{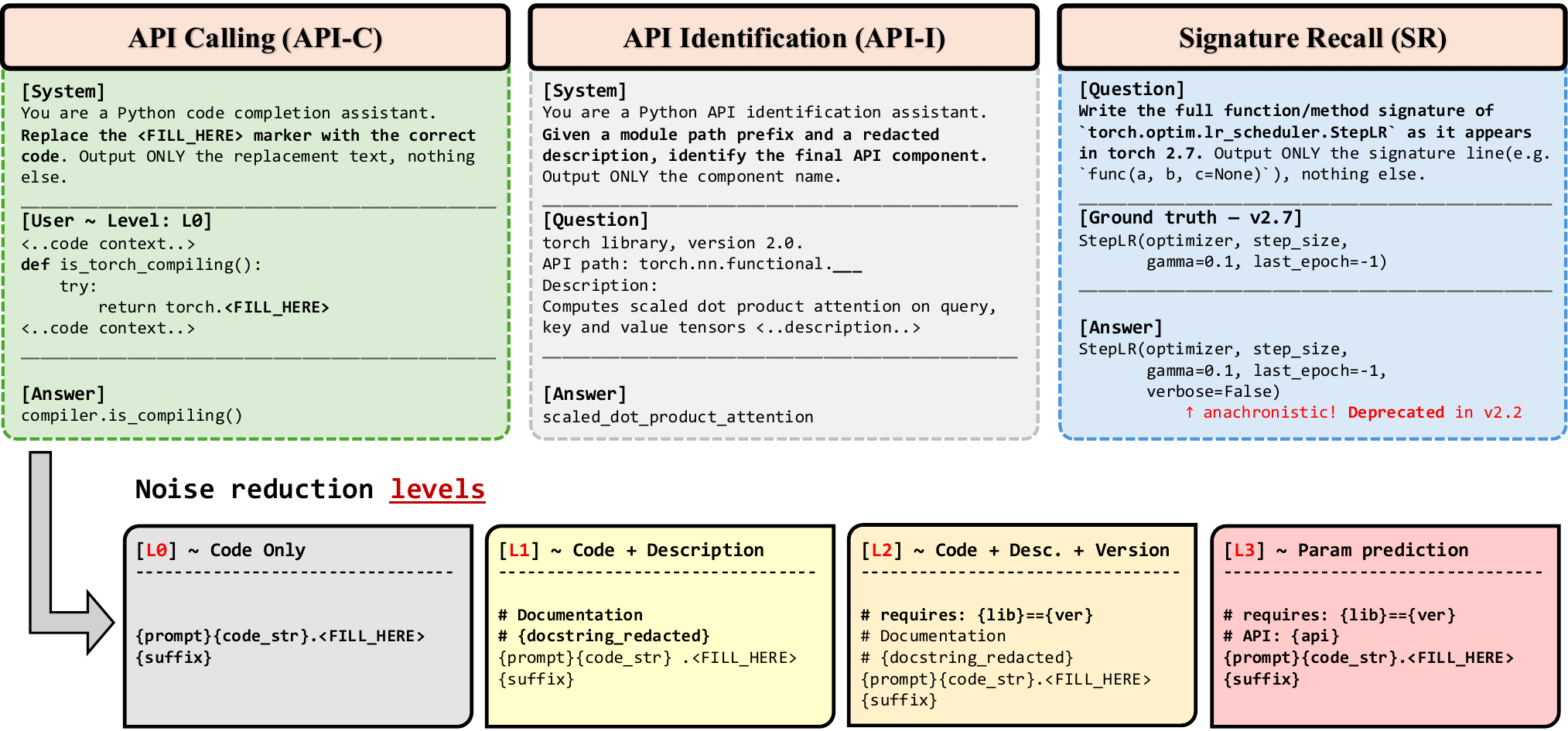}
  \caption{Overview of the three evaluation tasks and the \emph{API-C} noise reduction levels. \emph{API-C} uses real code contexts with progressively richer prompts (L0\textendash L3). \emph{API-I} isolates API identification from a redacted description. \emph{SR} probes signature recall from memory. Full prompt templates and response extraction details are in Appendix~\ref{app:prompts}.}
  \label{fig:tasks}
\end{figure*}

We present \namebenchmark{}, a benchmark that evaluates version-aware API knowledge through three complementary tasks, illustrated in Figure~\ref{fig:tasks}, each targeting a different modality of knowledge. 
All three tasks share a common foundation: API \emph{lifecycle analysis} that tracks the existence and signature of every public API across all tracked versions of a library. 
We first define each task (\S\ref{sec:tasks}), then describe the data collection pipeline (\S\ref{sec:data}), the lifecycle analysis and the hallucination taxonomy it enables (\S\ref{sec:taxonomy}), and the Software Evolution Understanding Score (\S\ref{sec:seus}).

\subsection{Evaluation Tasks}
\label{sec:tasks}

\paragraph{API Calling (API-C).}
Given an incomplete code snippet, the model must generate the missing API call.
This task evaluates model's knowledge of the library through real API calling scenarios.
Each sample originates from an open-source repository depending on a fixed library version.
Every API call has been extracted and verified using a static resolution engine, which maps each call site back to a specific entry point inside the pinned library version.
This traceability allows us to classify every model output according to a fine-grained hallucination taxonomy (\S\ref{sec:taxonomy}).

To obtain a cleaner signal from these samples, we introduce multiple \emph{noise reduction levels} that progressively reduce ambiguity and shift the diagnostic focus by manipulating the context provided to the model.
\textbf{L0} presents raw code only, measuring the model's ability to predict the correct API Call from context alone.
\textbf{L1} adds a redacted docstring that describes the target API without revealing its name, factoring out retrieval and isolating the model's ability to recall the correct API.
\textbf{L2} further adds an explicit version constraint (\texttt{\# requires:lib==ver}), testing whether the model adjusts its prediction when the target version is known and thereby exposing version-awareness gaps.
\textbf{L3} provides both the API name and the version but omits the signature, shifting the focus to fine-grained parameter recall across versions.

\paragraph{API Identification (API-I).}
Given the module path and a redacted function description, the model must identify the correct API name.
This task isolates \emph{declarative} API knowledge from the compositional reasoning required by realistic API calling.
Because instances of this task are derived directly from the documentation rather than from code samples, the available docstrings are filtered and preprocessed to ensure sufficient quality (see Appendix~\ref{app:doc-quality}), resulting in larger coverage of each library's versioned API surface compared to the API Calling task.

\paragraph{Signature Recall (SR).}

Given only the fully qualified API name and a target library version, the model must reproduce the complete function or method signature from memory, without code context, documentation or other hints.
This probes fine-grained parametric knowledge of argument names, defaults, and ordering, the aspects of an API that evolve most frequently across versions.

Each task uses the metric best suited to its output granularity.
\emph{API-C} and \emph{API-I} evaluate API selection and are scored by exact match (EM) on the predicted API name. 
\emph{API-C@L3} and \emph{SR} shift focus to parameter prediction, scored by recall and F1, respectively, against the ground-truth signature. 

\subsection{Data Collection}
\label{sec:data}

The benchmark construction pipeline is fully automated: starting from a crawl of public GitHub repositories, it joins real-world code with two independent views of each library's public API surface and produces a unified, version-indexed dataset. 

We decided to target three popular Python libraries, \texttt{PyTorch}, \texttt{NumPy}, and \texttt{SciPy}. 
Their rapid evolution and widespread adoption make them ideal candidates for version-aware evaluation, and together they provide over 125k evaluation instances spanning more than 16k APIs across 29 library versions. 
We describe the process of the library and version selection in more detail in Appendix~\ref{app:version-selection}, and share implementation details in Appendix~\ref{app:data-collection}.

\paragraph{Library Usage Collection}
For each target library, we mine public GitHub repositories whose \texttt{requirements.txt} file pins the library to a specific version. We keep only deduplicated repositories that use a permissive license and have received at least ten stars, a commonly adopted cutoff for excluding low-quality or abandoned projects~\cite{bogomolov_long_2024}.
Each repository is cloned at the manifest commit, and then, using static code analysis, we track imports, assignments, and alias propagation to identify every call site that reaches the target library. %
Each call site is further resolved to its fully qualified public API name by a static analysis engine running against the exact pinned library version, ensuring that type inference and module resolution operate against the same API surface the developer saw at commit time.

\paragraph{API surface: documentation and runtime.}
\label{sec:api-surfice-main}
We reconstruct the public API surface of each library version from two complementary sources.
The \textbf{documented surface} is obtained from the official Sphinx\footnote{\url{https://www.sphinx-doc.org}} \emph{objects.inv} inventory files, which library maintainers ship alongside their online documentation and which capture exactly the surface they intend to make public. 
However, these inventories are not always complete: widely used symbols can go undocumented for years. 
A notable finding is \texttt{torch.Size}, which was part of PyTorch's public API since its earliest releases, but did not appear in the official documentation until version~2.4 (July 2024).\footnote{\url{https://github.com/pytorch/pytorch/issues/12532}}
We complement the inventory with direct package introspection: for each library version, we install it into a dedicated environment, walk its module tree, and extract the full signature and canonical Python identity of every discovered function and method, forming the \textbf{runtime surface}.
An API is considered \emph{present} at a given version only when both sources agree: the documentation inventory confirms it as a public symbol and the runtime introspection engine successfully resolves it. An API previously confirmed but later dropped from documentation is treated as deprecated, regardless of whether its runtime symbol persists. 
This produces the compatibility matrices and a per-\emph{(API, version)} documentation dictionary that together form the backbone of \namebenchmark{}.

\smartparagraph{Compatibility matrices.}
The \emph{API compatibility matrix}
$\mathbf{A} \in \{0,1,\textsc{na}\}^{|\mathcal{F}|\times|\mathcal{V}|}$
records, for each API $f$ and version $v$, whether $f$ exists as a documented public symbol.
The \emph{signature compatibility matrix}
$\mathbf{S} \in \{0,1,\textsc{na}\}^{|\mathcal{P}|\times|\mathcal{V}|}$ operates at finer granularity, tracking individual \emph{(API, signature)} pairs across versions.
Cells are marked \textsc{na} when the two sources disagree irreconcilably and no confident judgement can be made. 
Across all matrices these account for only 3.4\% of cells (ranging from under 1\% on SciPy to around 5\% on PyTorch), 
so the ground truth is effectively non-ambiguous for the vast majority of the API surface.

Together, the matrices induce a natural classification of every tracked API into two groups. \emph{Stable} APIs appear with an identical signature across all covered versions. \emph{Evolving} APIs either modify their signature or are introduced/deprecated within the covered period. Across our three target libraries, evolving APIs account for 33-71\% of all tracked symbols, confirming that \namebenchmark{} is well populated with exactly the cases that matter most for version-aware evaluation.

\paragraph{Tasks construction and resulting dataset.}

The compatibility matrices drive the construction of all three task pools.
\emph{API Call} is built from the version-pinned code corpus, the raw pool from which we limit the number of samples for \emph{(API, version)} to control its size while maintaining high coverage.
The two declarative tasks instead are constructed directly from the Compatibility matrices. 
\emph{API Identification} enumerates API-version pairs with known existence status, yielding API identification queries over the tracked version history. 
\emph{Signature Recall} enumerates versioned signature instances, retaining each \emph{(API, version, signature)} case for which the compatibility matrix records a valid signature. 
We report the number of instances per task, the number of APIs covered, and overall coverage in Table~\ref{tab:benchmark-coverage}. More details on the sampling are presented in Appendix~\ref{app:data-collection}. To the best of our knowledge, this is the largest version-indexed evaluation of Python library APIs to date.

\begin{table}[t]
\centering
\caption{Evaluation dataset after sampling and filtering. \textbf{S/E}: Ratio of stable to evolving APIs in the evaluated set. \emph{$\Omega$}: Coverage of the library's public API surface across all tracked versions.}
\label{tab:benchmark-coverage}
\footnotesize
\resizebox{\columnwidth}{!}{%
\begin{tabular}{@{}l c  rrrr  rrrr  rrrr@{}}
\toprule
& & \multicolumn{4}{c}{\textbf{API-C}} & \multicolumn{4}{c}{\textbf{API-I}} & \multicolumn{4}{c}{\textbf{SR}} \\
\cmidrule(lr){3-6} \cmidrule(lr){7-10} \cmidrule(l){11-14}
\textbf{Library} & \textbf{Versions} & \textbf{Samples} & \textbf{APIs} & \textbf{S/E} & $\Omega$ & \textbf{Samples} & \textbf{APIs} & \textbf{S/E} & $\Omega$ & \textbf{Samples} & \textbf{APIs} & \textbf{S/E} & $\Omega$ \\
\midrule
PyTorch & 10 & 16{,}444 & 1{,}556 & 5.82 & 30\% & 17{,}036 & 2{,}372 & 0.94 & 46\% & 25{,}656 & 3{,}473 & 0.40 & 67\% \\
NumPy   & 10 &  9{,}275 &   948  & 5.54 & 37\% &  9{,}385 & 1{,}101 & 2.65 & 43\% & 11{,}386 & 1{,}354 & 0.66 & 53\% \\
SciPy   &  9 &  3{,}948 & 1{,}063 & 4.54 & 29\% & 13{,}244 & 1{,}815 & 1.14 & 50\% & 18{,}052 & 2{,}497 & 0.26 & 69\% \\
\midrule
\textbf{Total} & & \textbf{29{,}667} & \textbf{3{,}567} & & & \textbf{39{,}665} & \textbf{5{,}288} & & & \textbf{55{,}094} & \textbf{7{,}324} & & \\
\bottomrule
\end{tabular}
}
\end{table}

\subsection{Hallucination Taxonomy}
\label{sec:taxonomy}

We create a hallucination taxonomy for \namebenchmark{} with a novel deterministic classification from a version-indexed view of the API surface.
The compatibility matrices enable a hallucination taxonomy that classifies every prediction deterministically by matrix lookup, without recourse to manual coding.
The taxonomy operates at two granularities, API-level and parameter-level, that share a common structure but differ in what they reveal about model failure.

A prediction is \textbf{Correct} when it matches the ground truth at the target version,
and \textbf{Unknown} when the predicted symbol has never appeared in any observed version of the library.
Unknown predictions correspond to what prior work broadly labels hallucination: the model fabricates a symbol that does not exist. Yet, for the Signature Recall task such predictions may be valid, for example, when a parameter is handled in method implementation via \texttt{**kwargs}.

The two intermediate categories carry the diagnostic power of the taxonomy. 
An \textbf{Invalid} prediction, defined only at API granularity, retrieves a public symbol that genuinely exists at the target version but is not the intended one.
This category separates retrieval confusion from fabrication: the model stays within the library's actual surface but selects the wrong entry point, a failure of precision rather than of grounding. 
An \textbf{Anachronistic} prediction, defined at both granularities, is a symbol that exists somewhere in the library's history but not at the requested version. 
Anachronistic predictions isolate temporal confusion from general hallucination, the distinctive diagnostic that no prior classification scheme provides for individual predictions.

\subsection{Software Evolution Understanding Score}
\label{sec:seus}
To evaluate the multi-task, multi-version performance, we introduce the \textbf{Software Evolution Understanding Score} (SEUS), designed around three principles. 
First, a model must perform well on both stable and evolving APIs simultaneously. 
Second, performance must be consistent across versions: a \emph{stability penalty} discourages models that perform well on some versions but poorly on others. 
Third, we apply an explicit \emph{anachronism penalty} derived from hallucination taxonomy~(\S\ref{sec:taxonomy}).

We calculate SEUS as follows. Let \(S_v\) denote the score on \emph{stable} APIs, which are available and stable across versions, and let \(E_v\) denote the score on \emph{evolving} APIs. 
Both scores are averaged across the four task-level metrics: API-C@L2 exact match, API-C@L3 parameter recall, API-I exact match, and SR parameter F1. 
We combine $S_v$ and $E_v$ using their harmonic mean, $B_v = 2S_vE_v/(S_v+E_v)$.
This formulation ensures that the score is high only when the model performs well on both stable and evolving APIs; strong performance on stable APIs alone cannot mask weak version-sensitive performance. 
The final model score is the average across all $N = |\mathcal{L} \times \mathcal{T}|$ library--task pairs:
\[
\mathrm{SEUS} = \frac{1}{N} \sum_{(\ell,\, t)\, \in\, \mathcal{L} \times \mathcal{T}}
\left(
\mathbb{E}_v[B_v]
- \underbrace{\lambda \cdot \mathrm{Std}_v(E_v)}_{\text{stability penalty}}
- \underbrace{\gamma \cdot A_{\ell,t}}_{\text{anachronism penalty}}
\right)
\]
where $A_{\ell,t}$ is the anachronistic error rate on evolving APIs and $\lambda = \gamma = 0.5$. 
The stability penalty targets $E_v$ rather than $B_v$ because SEUS 
specifically aims to penalize inconsistent version-specific knowledge. 
Since stable APIs are identical across versions, only variation in 
$E_v$ reflects whether a model has truly learned to distinguish between versions.
The anachronism penalty addresses a distinct failure mode: 
predictions that are not merely incorrect but correspond to a 
real API from a different version of the same library. 
Unlike generic errors, these reveal that the model actively confuses versions rather 
than lacking knowledge entirely, making them a direct indicator of poor version awareness.
We verify that model rankings are robust to the choice of 
$\lambda$ and $\gamma$ in Appendix~\ref{app:seus-sensitivity}.

%% file: sections/122-experiments.tex
\section{Evaluation}
\label{sec:results}
\smartparagraph{Models.} We evaluate three frontier model families across multiple generations: OpenAI's GPT-4.1, GPT-5, GPT-5.1, GPT-5.4, and GPT-5.5~\cite{noauthor_introducing_2026};
Google's Gemini 2.0 Flash, Gemini 2.5 Flash, and Gemini 3 Flash~\cite{team_gemini_2025};
and Anthropic's Claude Sonnet 4 and Sonnet 4.6~\cite{noauthor_introducing_nodate}. 
To complement this evaluation with a controlled analysis on models' scaling, we additionally include Qwen3.5~\cite{team_qwen35-omni_2026} at three sizes (35B, 122B, and 397B), isolating the effect of model size within a single model family while keeping architecture and training procedure fixed.
All models are queried using zero temperature where supported and no thinking budget.

\subsection{Model Performance on Evolving APIs}

\begin{figure}
  \centering
      \includegraphics[width=1\linewidth]{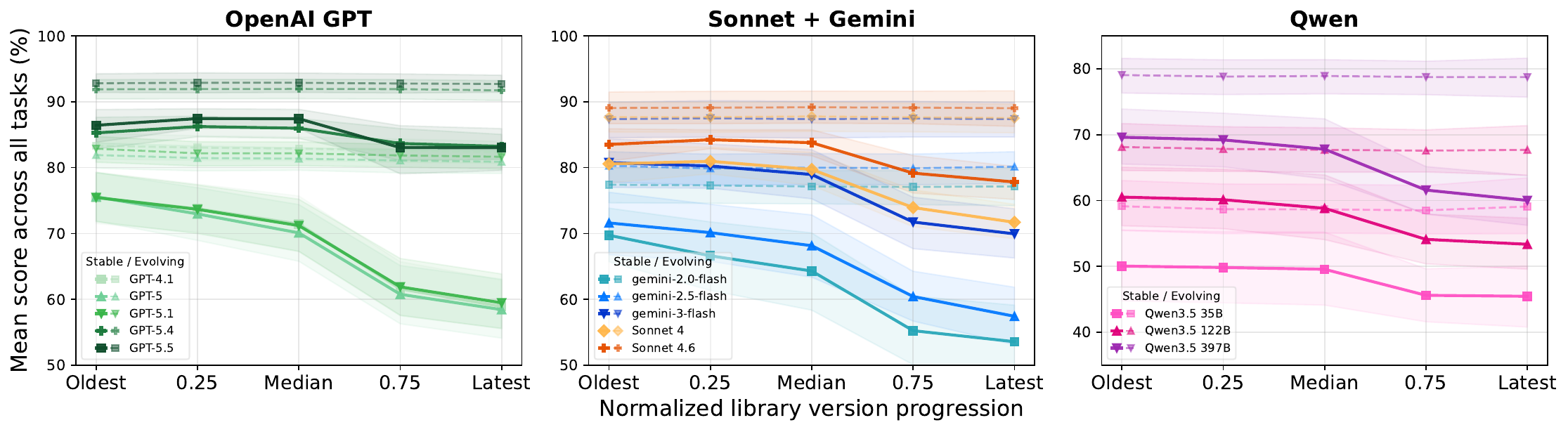}  
  \caption{Mean score across \emph{API-C@L1}, \emph{API-I}, and \emph{SR} as a function of normalized library version progression. Stable APIs (dashed) remain flat; Evolving APIs (solid) degrade toward recent versions across all model families. Further breakdown in Appendix~\ref{app:app-results}.}
  \label{fig:families_row_all_norm}
\end{figure}

Figure~\ref{fig:families_row_all_norm} aggregates the mean score across all tasks and maps each library's release history onto a unified progression from oldest to latest evaluated version.
The results reveal a clear pattern: the degradation is not uniform between versions. 
Performance on Evolving APIs is closer to Stable APIs for older releases, but erodes progressively towards the most recent ones, with the largest gaps observed for APIs from the latest versions.
Newer model generations partially recover this gap: GPT-5.5 maintains the narrowest spread between Stable and Evolving APIs throughout the entire progression, suggesting that scale and more recent pre-training data push the knowledge boundary forward. 
Within the Qwen3.5 family, model size shifts quality upward but leaves the decay curve unchanged. 
All three Qwen3.5 variants degrade in near-parallel steps when moving between library versions, confirming that capability scales with model size but the temporal pattern does not. 
Appendix~\ref{app:cutoff} shows that this decay is not explained by proximity to the models' knowledge cutoffs. 

\begin{figure}
  \centering
  \includegraphics[width=1\linewidth]{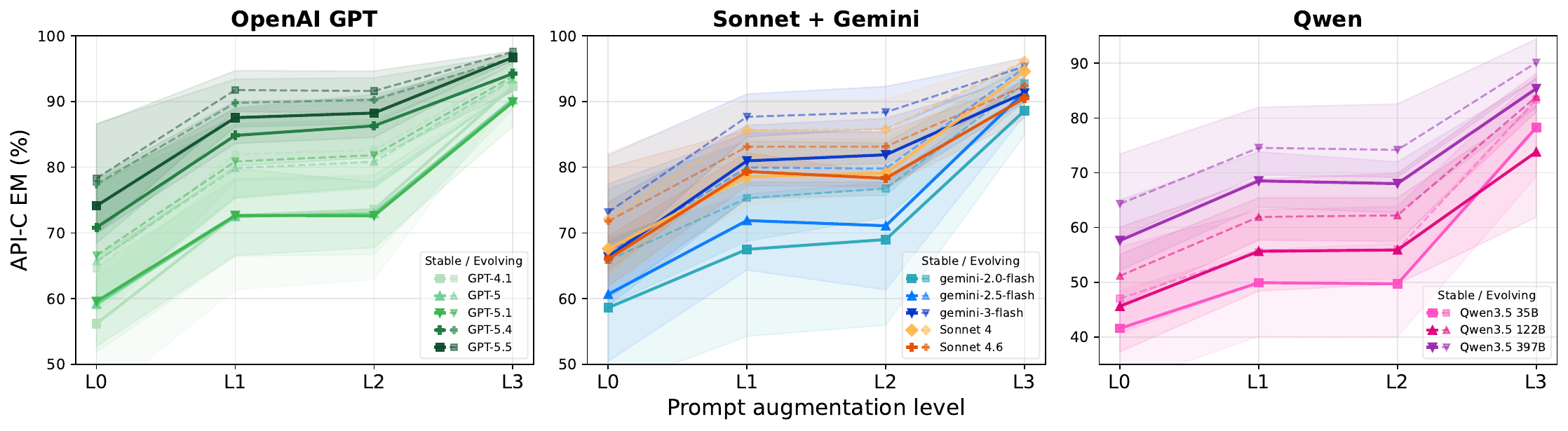}
  \caption{API-C noise reduction staircase aggregated over libraries. Redacted documentation context drives the L0$\to$L1 gain (code$\to$code+doc); adding a version constraint at L2 (code+doc+version) yields no further improvement. L3 (API name+version) shifts the task to parameter prediction.}
  \label{fig:staircase_torch}
\end{figure}

The performance deficit on Evolving APIs is remarkably consistent across all four model families, all three libraries, and all three tasks: even GPT-5.5 and Sonnet 4.6 lose 5-10 points on Evolving APIs despite their strong Stable baselines, confirming that the degradation is a structural property rather than a capability ceiling. 
The effect also replicates on three additional libraries (Appendix~\ref{app:replication}). 

As shown in Figure~\ref{fig:staircase_torch}, context enhancement in the API Calling task results in a consistent 10-20 point gain when adding description of the target method (L0$\rightarrow$L1) across all models, confirming that models perform genuine knowledge retrieval rather than memorized code matching. 
In contrast, providing version information (L1$\rightarrow$L2) does not yield gains: models do attempt to adjust their predictions when given an explicit version constraint, yet accuracy remains close, revealing that the version tag triggers a response without the underlying version-specific knowledge to back it up.  

\begin{figure}
  \centering
  \includegraphics[width=1\linewidth]{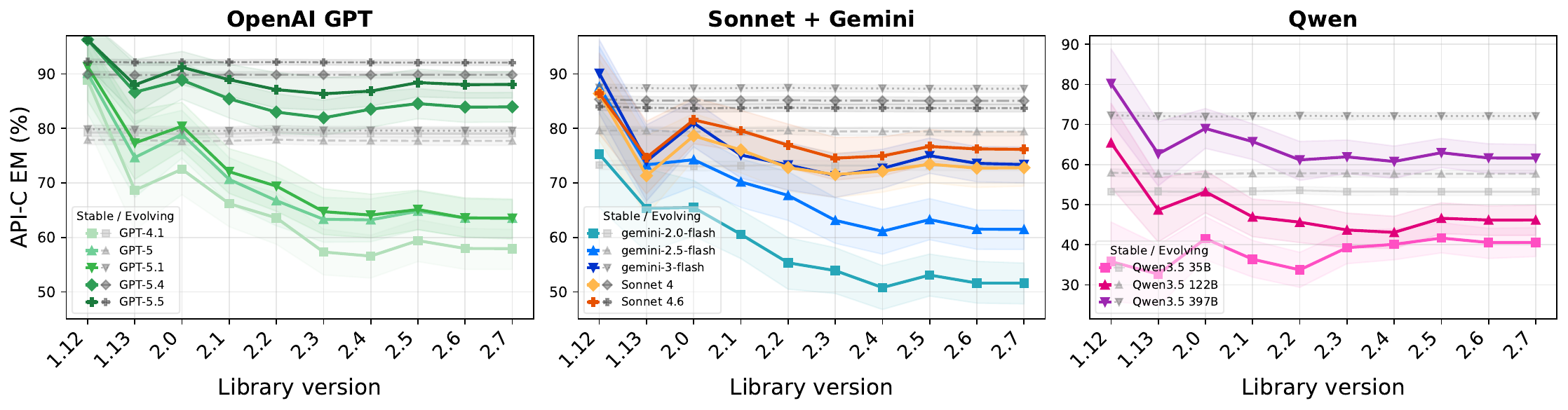}  
  \caption{\emph{API-C@L1} exact match on Evolving APIs per PyTorch release. Per-version resolution on real code completion instances across all four model families.}
  \label{fig:families_row_lifecycle_torch}
\end{figure}

Figure~\ref{fig:families_row_lifecycle_torch} reports API Calling scores per PyTorch release, grounding the aggregate trend in real-world examples of API calls. The per-version resolution on real usage samples is a distinctive property of \namebenchmark{}: degradation is not just detectable in the aggregate but also localizable to individual releases.

\subsection{Hallucination Taxonomy}
\begin{figure}
  \centering
  \includegraphics[width=1\linewidth]{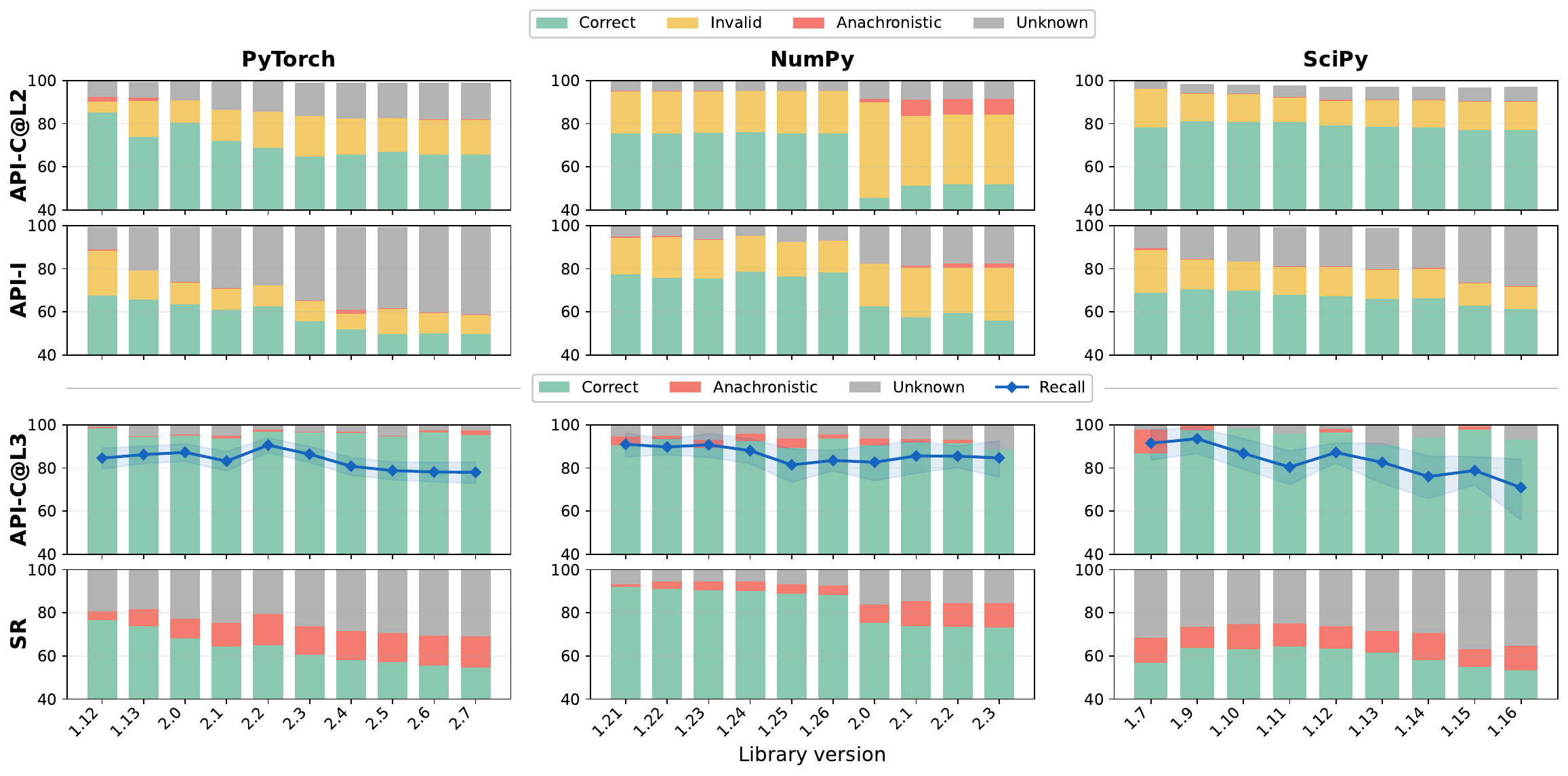}  
  \caption{Per-version taxonomy breakdown for \textbf{GPT-5.1} across API-level (API Call with version \emph{API-C@L2}, API Identification \emph{API-I}) and parameter-level (API Call parameter recall \emph{API-C@L3}, Signature Recall \emph{SR}) predictions. The blue line reports parameter recall for \emph{API-C@L3}.}
  \label{fig:taxonomy-gpt-5.1}
\end{figure}

The taxonomy introduced in \S\ref{sec:taxonomy} allows us to classify every error
deterministically by matrix lookup. Figure~\ref{fig:taxonomy-gpt-5.1} reports the full
breakdown per version and task level for GPT-5.1 across libraries.

\smartparagraph{API-level taxonomy.}
Across the API selection tasks (API-C@L2 and API-I), errors cluster into two dominant categories:
\emph{Invalid} predictions, where the model retrieves a different documented public symbol at the target version rather than the ground-truth one, 
and \emph{Unknown} predictions, where the model produces a symbol not observed anywhere in the API history of the library (Table~\ref{tab:taxonomy-apis}). 
This balance changes depending on model generation and scale. In line with previous observations, successive GPT generations show a consistent reduction in both error categories, with hallucination rates declining most sharply.
Looking instead at parameter scale variation across the Qwen3.5 family, we observe a different pattern: both \emph{Unknown} and \emph{Invalid} rates remain substantial even at 397B, indicating that an increased scale does not compensate for the underlying knowledge deficiency.
At the API selection level, \emph{Anachronistic} predictions remain consistently negligible across all models and versions, yet this should not be mistaken for genuine version awareness. When models fail here, they systematically fall back on high-frequency symbols that are valid across most versions, effectively concealing their lack of version-specific knowledge.

\smartparagraph{Parameter-level taxonomy.}
At the parameter level, anachronistic errors constitute a substantial and consistent fraction of the error budget across all libraries and models. \emph{Anachronistic} rates on Signature Recall (\emph{SR}) generally range from 6-14\%, with \emph{Unknown} parameters reaching 7-42\% depending on the library and model (Table~\ref{tab:taxonomy-params}). Figure~\ref{fig:taxonomy-gpt-5.1} further shows that this anachronistic pressure intensifies with recency: as the target version increases, the correct fraction shrinks while anachronistic predictions accumulate. 
Signature Recall (\emph{SR}) consistently amplifies both error types relative to the code generation scenario (\emph{API-C@L3}) across all libraries. 
This is not because models hallucinate more in isolation, but because explicitly enumerating all parameters forces the model to commit to version-specific choices, surfacing latent errors that remain concealed when only a subset is generated. 
In \emph{API-C@L3}, degradation manifests primarily as reduced \emph{parameter recall} rather than increased hallucination, suggesting models implicitly hedge by emitting only their highest-confidence parameters. These trends are consistent across model families, with later GPT generations and larger \emph{Qwen3.5} models showing only incremental gains. 

\smartparagraph{SEUS Leaderboard.}
Table~\ref{tab:seus-leaderboard} ranks the best model from each family by SEUS (see the full leaderboard in Appendix~\ref{app:seus}). 
GPT-5.4 leads with a SEUS of 86.0, combining the highest evolving-API score of 86.4 with the lowest stability penalty of 2.0\,pp, 
indicating both strong and consistent version-specific knowledge.
Notably, GPT-5.5 achieves a slightly higher stable score of 92.3 compared to GPT-5.4's 92.0, but falls behind due to 
greater version-to-version inconsistency, with a higher stability penalty of 2.7\,pp.
This is a case where SEUS rewards consistency over raw performance.

Three tiers emerge: a top tier above 80 (GPT-5.4, GPT-5.5, Sonnet~4.6), 
a middle cluster between 70 and 78 spanning four model families, and a lower tier below 70. 
Within the middle tier, the stability penalty becomes the primary differentiator.
Sonnet~4 and Gemini~3 Flash, for instance, share the same stable and evolving scores, 
yet they are separated by their penalty profiles.

We also report, as a standalone diagnostic outside of SEUS, retention: the fraction of stable-API performance a model preserves on evolving APIs.
It isolates how much performance a  model loses specifically due to API evolution: 
top-tier models preserve over 93\%, while mid-tier models drop below 86\%. 
However, retention is uninformative for less powerful models---Qwen3.5~122B 
achieves 87.8\% retention despite ranking 12$^{th}$, because both 
its stable and evolving scores are low. The anachronism penalty 
tells a similar story, growing from 0.9\,pp at the top to 
1.8\,pp at the bottom: less powerful models are not only less accurate 
on evolving APIs but also more prone to confusing one version 
with another.

\begin{table}[t]
\centering
\caption{Model leaderboard by SEUS for the best model of each family (full table in Appendix~\ref{app:seus}). Stable and Evolving are the average scores on stable and evolving APIs; Retention is Evolving/Stable (\%). The penalties measure version instability and anachronistic (temporal-confusion) predictions.}
\label{tab:seus-leaderboard}
\footnotesize
\setlength{\tabcolsep}{2.2pt}
\begin{tabular}{@{}r l c c c  c c c }
\toprule
\textbf{Rank} & \textbf{Model} & \textbf{SEUS} & \multicolumn{3}{c}{\textbf{Performance}} & \multicolumn{2}{c}{\textbf{Penalty}}\\
\cmidrule(lr){4-6}\cmidrule(lr){7-8}
& & & \textbf{Stable} & \textbf{Evolving} & \textbf{Retention} & \textbf{Stability} & \textbf{Anachronism} \\
\midrule
1 & GPT-5.4 & \textbf{86.0} & 92.0 & 86.4 & 94.1 & 2.0 & 1.0 \\
2 & GPT-5.5 & 85.1 & 92.3 & 86.1 & 93.5 & 2.7 & 0.9 \\
3 & Sonnet 4.6 & 81.0 & 89.3 & 82.4 & 92.4 & 3.4 & 1.0 \\
5 & Gemini 3 Flash & 77.3 & 87.8 & 78.5 & 89.4 & 3.9 & 1.2 \\
10 & Qwen3.5 397B & 67.9 & 80.2 & 68.9 & 85.9 & 3.9 & 1.4 \\
\bottomrule
\end{tabular}
\end{table}

%% file: sections/222-related.tex
\section{Related Work}

\smartparagraph{Version-aware code generation and continuous learning.} 
Although several benchmarks evaluate version-specific code generation, they generally isolate the evaluation from the broader continuous learning challenge of library evolution. 
Execution-based benchmarks like GitChameleon~\cite{misra_gitchameleon_2025} and CodeUpdateArena~\cite{liu_codeupdatearena_2024} offer high-fidelity evaluation but are inherently limited by manual curation, restricting their scope to a few hundred isolated breaking changes. 
Conversely, broader datasets such as VersiCode~\cite{wu_versicode_2024} rely on heuristic string-matching, which fails to establish the deterministic, per-API ground truth necessary to accurately evaluate temporal knowledge. 
LibEvolutionEval~\cite{kuhar_libevolutioneval_2024} explicitly targets library evolution, but its evaluation is dominated by a narrow surface of unchanged version-agnostic APIs. Furthermore, its coarse F1 over identifiers metric conflates version-specific errors with general hallucinations, limiting its diagnostic power.
Version numbers alone do not identify breaking changes: under strict semantic versioning, minor and patch releases introduce no breaking changes, but adherence is not a given~\cite{ochoa2022breaking}. Our compatibility matrices therefore record breaking changes per method, independently of what the version number signals.

\smartparagraph{Hallucination taxonomies and temporal knowledge.}
Existing taxonomies of code-generation hallucinations~\cite{liu_beyond_2026, tian_codehalu_2025, zhang_llm_2024,
spracklen_we_2025, jain_mitigating_2024, chen_towards_2025} are constructed through manual analysis of model outputs 
and organise errors into thematically defined buckets: intent, context, or knowledge
conflicts~\cite{liu_beyond_2026, zhang_llm_2024}, execution-based failure modes
such as mapping, naming, resource, and logic errors~\cite{tian_codehalu_2025}, 
or single phenomena such as non-existent packages~\cite{spracklen_we_2025}. 
Closer to our setting, ~\citet{wang_llms_2025} quantify deprecated API usage in code completion, 
and LibEvolutionEval~\cite{kuhar_libevolutioneval_2024} classifies APIs by lifecycle stage (introduced, modified, deprecated, unchanged); 
neither, however, provides a taxonomy that, for an individual prediction, separates temporal confusion from outright fabrication. 
\namebenchmark{} addresses this gap by introducing a high-precision taxonomy driven entirely by automated lookup against
version-indexed API compatibility matrices, deterministically classifying errors 
at both the API and parameter levels.

The broader NLP literature has established that LLMs struggle with evolving factual knowledge~\cite{zhu_evolvebench_2025, wei_time_2025}, and prior software engineering studies have quantified deprecated API usage~\cite{wang_llms_2025}. Our precise taxonomy empirically grounds this continuous learning failure in code generation at scale and with high-precision classification. 
Ultimately, \namebenchmark{} exposes the next structural bottleneck in the evolution of AI for software engineering: to move beyond static code generation, models must master the continuous, version-differentiated evolution of real-world software ecosystems.

%% file: sections/444-conclusion.tex
\section{Limitations}
\label{sec:limitations}
All models are evaluated with zero thinking budget, a deliberate choice to isolate parametric knowledge rather than reasoning capability. This is the same knowledge regime that underlies API selection in interactive and agentic coding, where calls are composed implicitly rather than deliberated over. To test this choice, we repeated the PyTorch evaluation with extended reasoning on GPT-5.1, Gemini 3 Flash and Sonnet 4.6 (Appendix~\ref{app:reasoning}). Reasoning raises absolute scores, but the gap between stable and evolving APIs still widens at the same rate across releases, and the curves keep their shape on every task: a larger thinking budget does not recover the version-specific facts collapsed during training, which is the failure mode we target.

The \textbf{API-C }task is mined from public repositories that may overlap with model training corpora. Direct contamination probes are infeasible for the closed-weight models we evaluate. We rely on three indirect signals that together argue against memorization as the dominant driver: (i) memorization would produce flat API-C scores across versions, whereas we observe systematic version-correlated degradation; (ii) contamination would inflate API-C scores relative to the declarative tasks, which are constructed from compatibility matrices rather than mined code; instead, we find consistent degradation patterns across all three tasks; and (iii) the large \textbf{L0}$\rightarrow$\textbf{L1} gain from adding a redacted docstring demonstrates that models rely on semantic cues rather than verbatim recall. That said, contamination cannot be fully ruled out, and we encourage future work to develop tighter contamination-control protocols for version-indexed benchmarks.

Evaluating specific agentic harnesses end to end, where retrieval and tests are available, is valuable future work; our results characterize the knowledge such harnesses build on. A related but distinct problem is variation between simultaneous implementations, such as SQL dialects. \namebenchmark{} targets variation within a single artifact over time, where the training corpus blends points of its history; extending the evaluation to cross-implementation variation is adjacent future work.

\section{Conclusion}

We introduced \textbf{\namebenchmark{}}, a multi-task benchmark that probes version-aware API
knowledge at a scale and precision beyond existing evaluations, and the Software
Evolution Understanding Score (SEUS), a versioned composite metric designed to
track progress on this problem as both models and libraries evolve. 
Our evaluation across four model families and thirteen models establishes a clear
empirical picture: current LLMs are fundamentally \textit{version-oblivious}. They benefit
substantially from the documentation context, but gain nothing from an explicit
version constraint. This indicates that the version tag activates a retrieval cue
with no version-differentiated knowledge behind it. 
The resulting degradation on evolving APIs is structural: invariant to model scale within a family, narrowing
but persisting across model generations, and driven by temporal confusion rather
than fabrication. 
These findings carry concrete implications for both researchers and practitioners. They reveal that temporal knowledge stratification in code generation is a persistent, first-class capability gap, not an artifact of benchmark sensitivity or prompt framing, that neither scale nor training recency has yet proven sufficient to close.

%% file: sections/999-appendix.tex
\input{sections/appendix/versions_selection}

\clearpage
\input{sections/appendix/results}
\clearpage
\input{sections/appendix/seus}

\clearpage
\input{sections/appendix/data_collection}

\clearpage
\input{sections/appendix/doc_quality}

\clearpage
\input{sections/appendix/prompts}
\clearpage
\input{sections/appendix/examples}

%% file: sections/appendix/versions_selection.tex
\section{Library-Version Selection}
\label{app:version-selection}

A core design goal of \namebenchmark{} is to evaluate models on the libraries and versions that developers actually use, not on a curated wishlist. 
We ground both the library selection and the version coverage in a large-scale, empirical analysis of real
dependency manifests.

\smartparagraph{Library selection.}
PyPI download statistics\footnote{\url{https://pypistats.org/top}} conflate end-user installs with CI runs, mirror traffic, and transitive dependencies, and therefore poorly reflect what developers actually \emph{import} in their code.
To obtain a more faithful signal, we crawl public GitHub repositories and parse roughly 70k \texttt{requirements.txt} files, applying the same repository filters used in the main pipeline (permissive license, at least ten stars, deduplication).

\begin{figure}[h]
\centering
\includegraphics[width=0.85\linewidth]{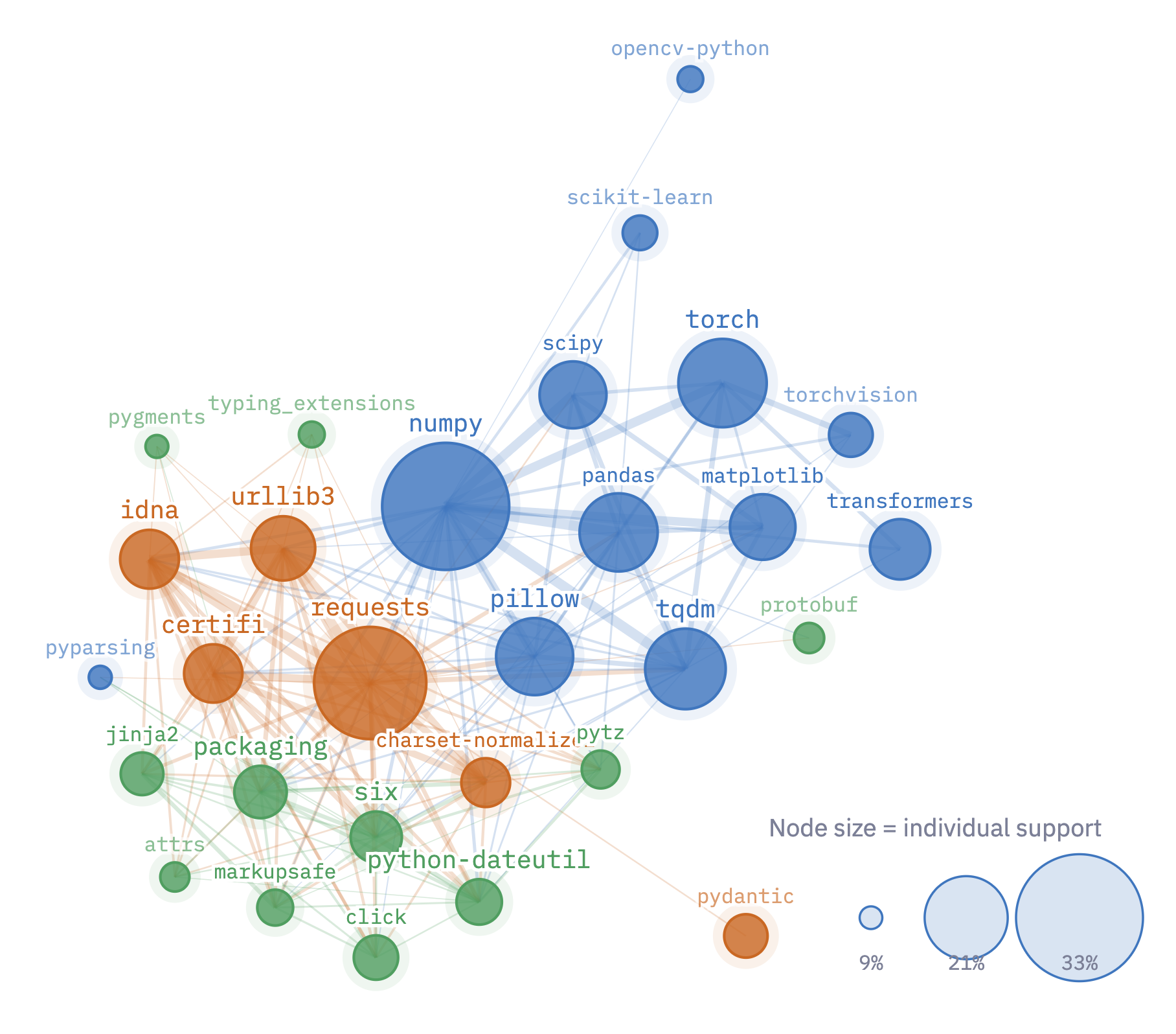}
\caption{Library co-occurrence graph mined from $\sim$70k GitHub \texttt{requirements.txt} files.
Node size reflects individual support; edge weight reflects joint support. Filters:
$\geq 8\%$ individual, $\geq 7\%$ joint.}
\label{fig:lib-graph}
\end{figure}

Each manifest is treated as a vote for the libraries it pins, yielding a co-occurrence graph in which node size reflects individual support (the fraction of repositories listing the library) and edge weight reflects joint support.
Figure~\ref{fig:lib-graph} visualizes the resulting graph, restricted to nodes with individual support $\geq 8\%$ and edges with joint support $\geq 7\%$. Three clusters emerge naturally: a scientific-Python core (blue), a web/HTTP cluster (orange), and a tooling/serialization cluster (green). 
From the scientific cluster, which dominates by both node mass and inter-library co-occurrence, we select PyTorch, NumPy, and SciPy as the target libraries for the benchmark.
They are the most heavily used scientific libraries with non-trivial API evolution histories, and together they exercise both deep-learning and general numerical workloads.

\begin{figure}[h]
\centering
\includegraphics[width=0.7\linewidth]{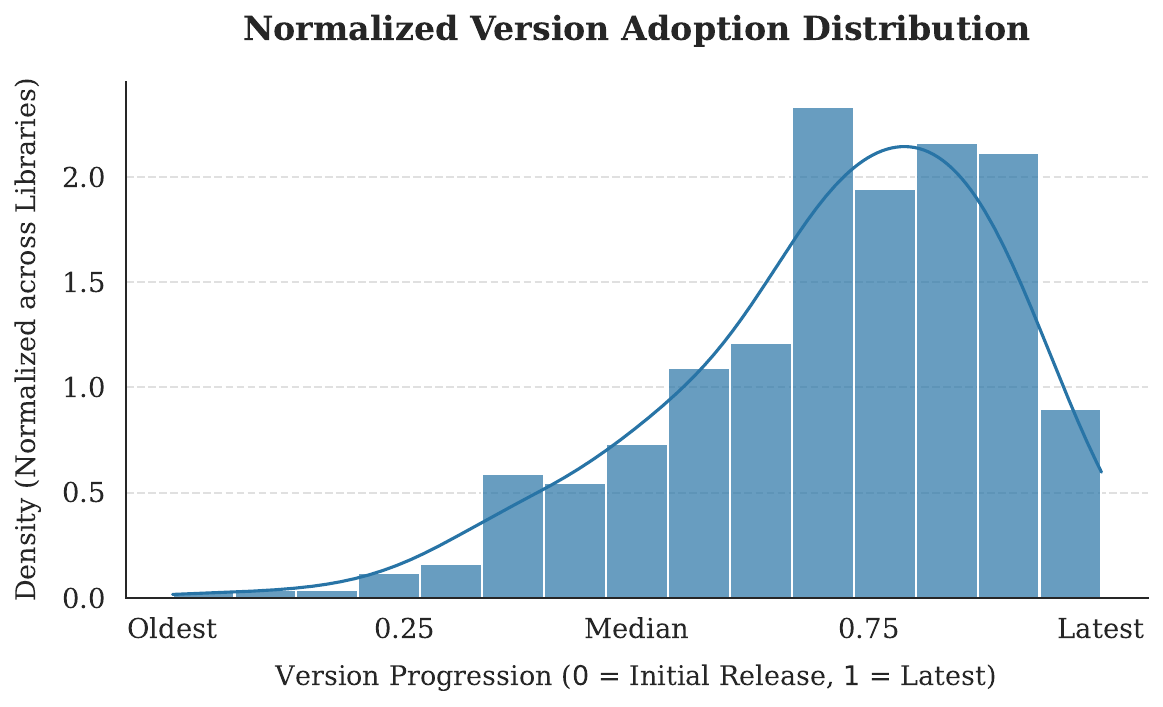}
\caption{Normalized version-usage density aggregated across all libraries mined from GitHub
manifests. The horizontal axis maps each library's release history onto $[0,1]$.}
\label{fig:version-trends}
\end{figure}

\smartparagraph{Version selection.}
For each selected library, we extract every pinned version specifier (\texttt{==} operator) from
the same manifest corpus, yielding a real-world frequency distribution over library versions
weighted by how often each version appears in production-style dependency files rather than by
release recency. Aggregating these distributions across all mined libraries already reveals a
clear pattern: Figure~\ref{fig:version-trends} normalizes each library's release history onto
the unit interval ($0 =$ oldest pinned release, $1 =$ latest) and shows a broad, right-skewed
density with substantial mass well below the midpoint. Across the ecosystem, a large fraction
of pinned dependencies trail their latest release by a non-trivial margin, which is precisely
what makes version-aware evaluation a meaningful problem.

\begin{figure}[h]
\centering
\begin{subfigure}[t]{0.45\linewidth}
  \includegraphics[width=\linewidth]{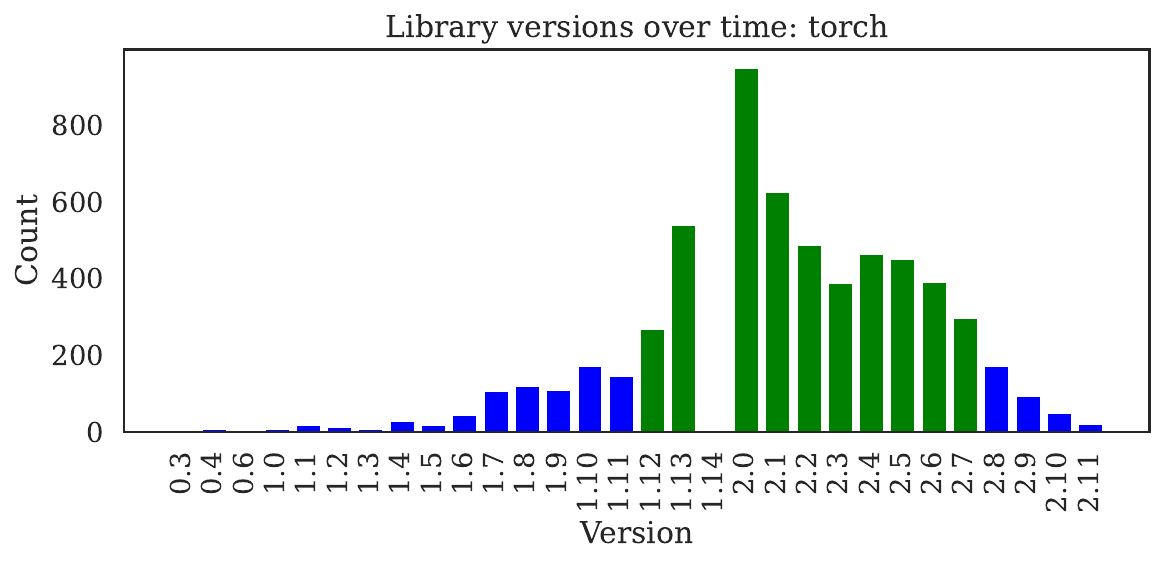}
  \caption{PyTorch}
\end{subfigure}
\hfill
\begin{subfigure}[t]{0.45\linewidth}
  \includegraphics[width=\linewidth]{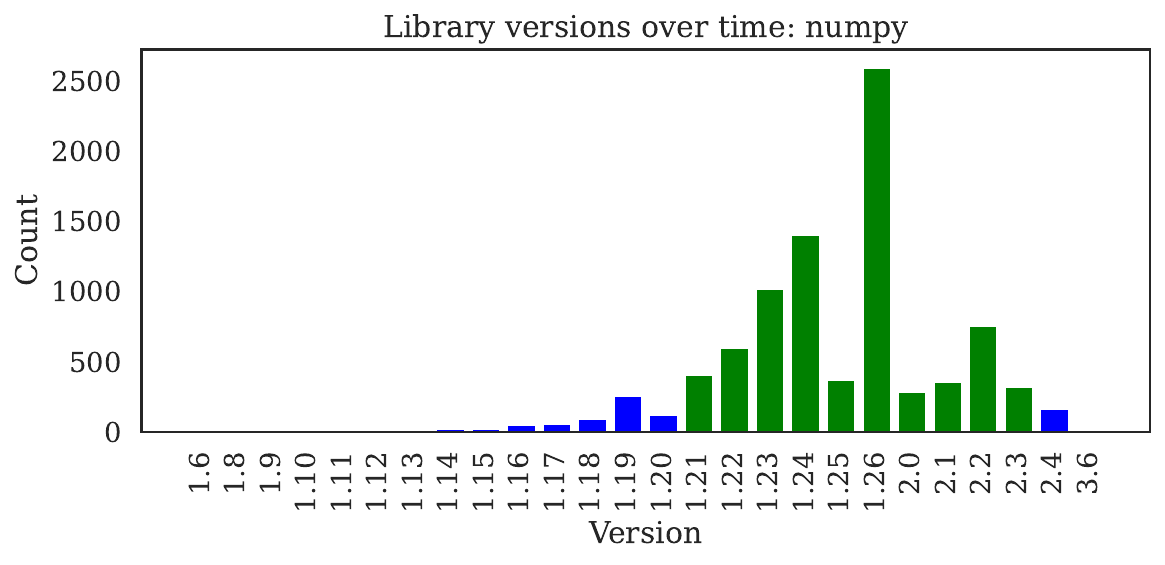}
  \caption{NumPy}
\end{subfigure}
\hfill
\begin{subfigure}[t]{0.45\linewidth}
  \includegraphics[width=\linewidth]{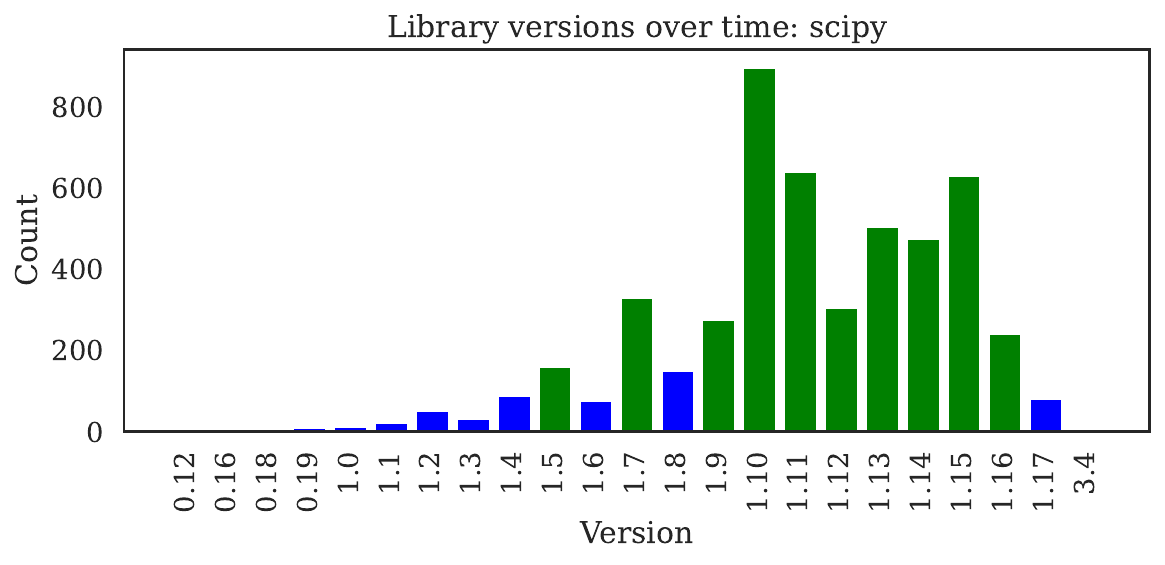}
  \caption{SciPy}
\end{subfigure}
\caption{Per-library version-usage distribution mined from pinned 
\texttt{==} specifiers in real
manifest files.}
\label{fig:lib-versions}
\end{figure}

The same pattern holds for our three target libraries individually. Figure~\ref{fig:lib-versions}
shows the per-library distributions: usage spans nearly the entire release history, with
substantial mass on versions several years old. We retain the top-10 most-used versions per
library that also reach a minimum support threshold of 100 repositories. For PyTorch and NumPy
this yields 10 covered versions each; for SciPy, version 1.5 is too old for our extraction toolchain (both documentation-inventory parsing and runtime introspection failed) and was excluded, leaving 9 covered versions.

%% file: sections/appendix/results.tex
\section{Expanded Results}
\label{app:app-results}

\subsection{Infrastructure}
\label{app:infra}
All commercial models were evaluated through their respective public API endpoints.
Open-source models, specifically the Qwen3.5 family at 35B, 122B, and 397B parameters,
were served on a dedicated cluster of 8 NVIDIA H200 GPUs using vLLM\footnote{\url{https://vllm.ai/}} for inference.
All evaluations used greedy decoding with zero temperature and no thinking budget, except for the reasoning ablation in Appendix~\ref{app:reasoning}.

\subsection{Results Breakdown}
\begin{figure}[h]
\centering
\includegraphics[width=1\linewidth]{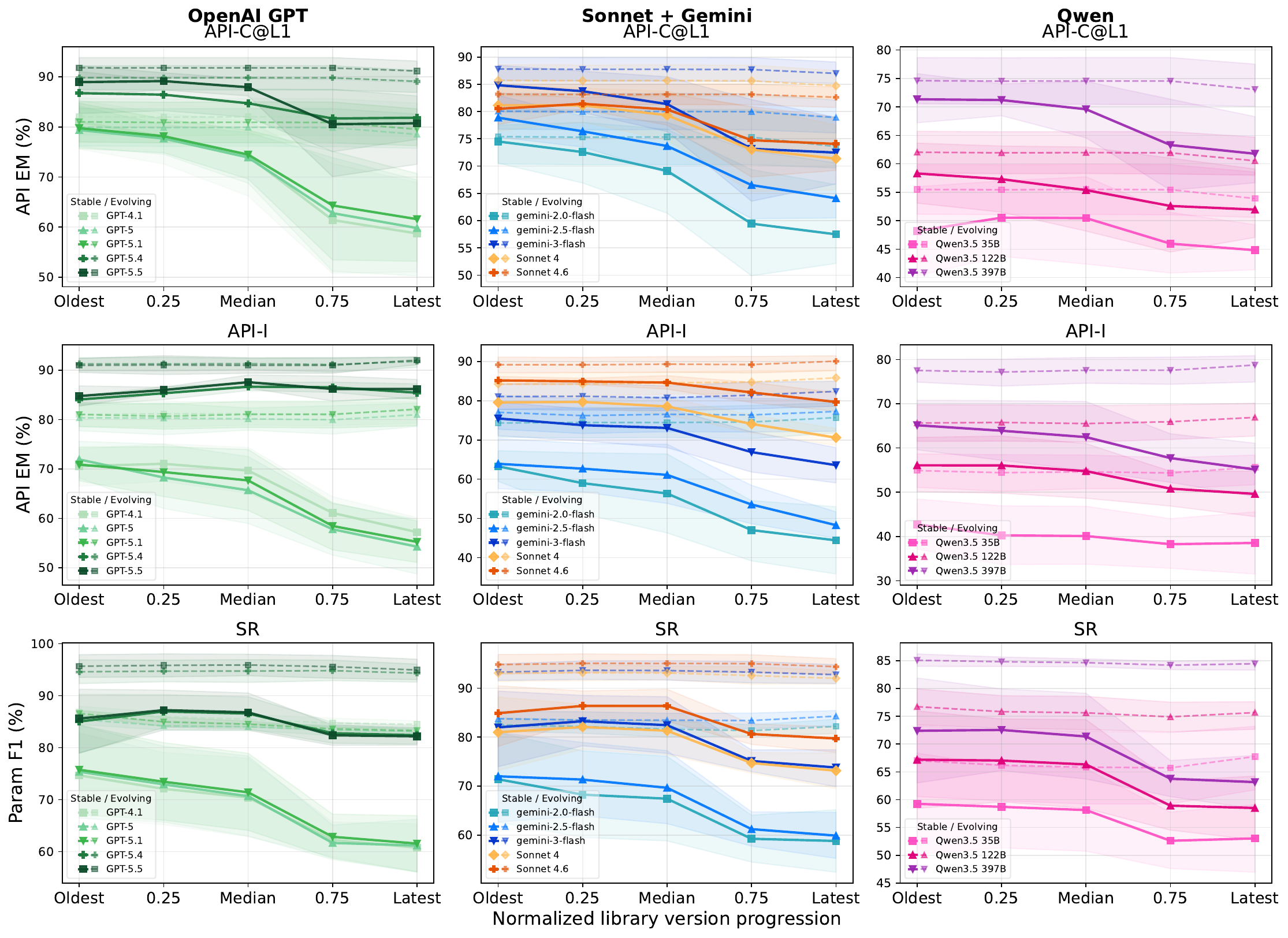}
\caption{Comprehensive performance breakdown across tasks, aggregated across libraries.}
\label{fig:families_row_all_norm_decomposition}
\end{figure}

This appendix provides a full breakdown of the results summarized in the main text. Figure~\ref{fig:families_row_all_norm_decomposition} decomposes the aggregate score of Figure~2 by task, reporting API-C@L1 API exact match, API-I API exact match, and SR parameter F1 separately for each model family. The task-level decomposition confirms that the degradation pattern on Evolving APIs is not driven by a single task: the Stable/Evolving gap and its progressive widening toward recent versions are consistently observed across all three evaluation modalities.

Table~\ref{tab:results-ci} reports the full per-model, per-library, per-task results with 95\% bootstrap confidence intervals, stratified by API stability category. Each cell reports the mean score on Stable and Evolving APIs separately, allowing direct comparison of version-sensitive degradation across models and libraries. The Stable/Evolving gap is statistically significant across all model families and libraries, with non-overlapping confidence intervals in the large majority of cases. Notably, the gap is most pronounced on PyTorch, which has the largest fraction of Evolving APIs and the longest covered version history among the three target libraries.

\input{sections/tables/results_appendix/results_with_ci}

\clearpage
\subsection{Taxonomy}

\begin{figure}[h]
  \centering
  \includegraphics[width=1\linewidth]{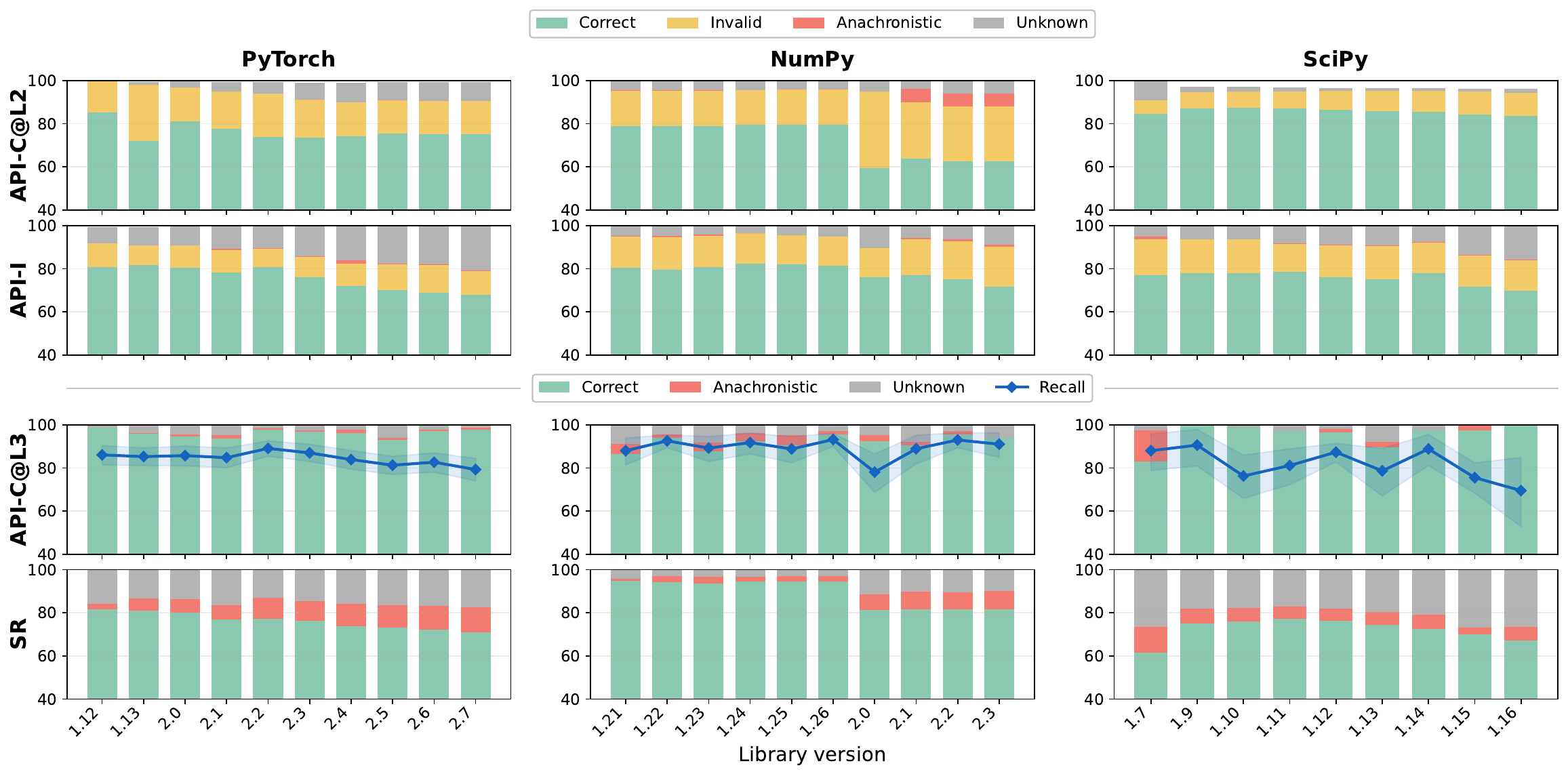}  
  \caption{Per-version taxonomy breakdown for \textbf{Sonnet 4} across API-level (API Call with version \emph{API-C@L2}, API Identification \emph{API-I}) and parameter-level (API Call parameter recall \emph{API-C@L3}, Signature Recall \emph{SR}) predictions. The blue line reports parameter recall for \emph{API-C@L3}.}
  \label{fig:taxonomy-sonnet-4}
\end{figure}

\begin{figure}[h]
  \centering
  \includegraphics[width=1\linewidth]{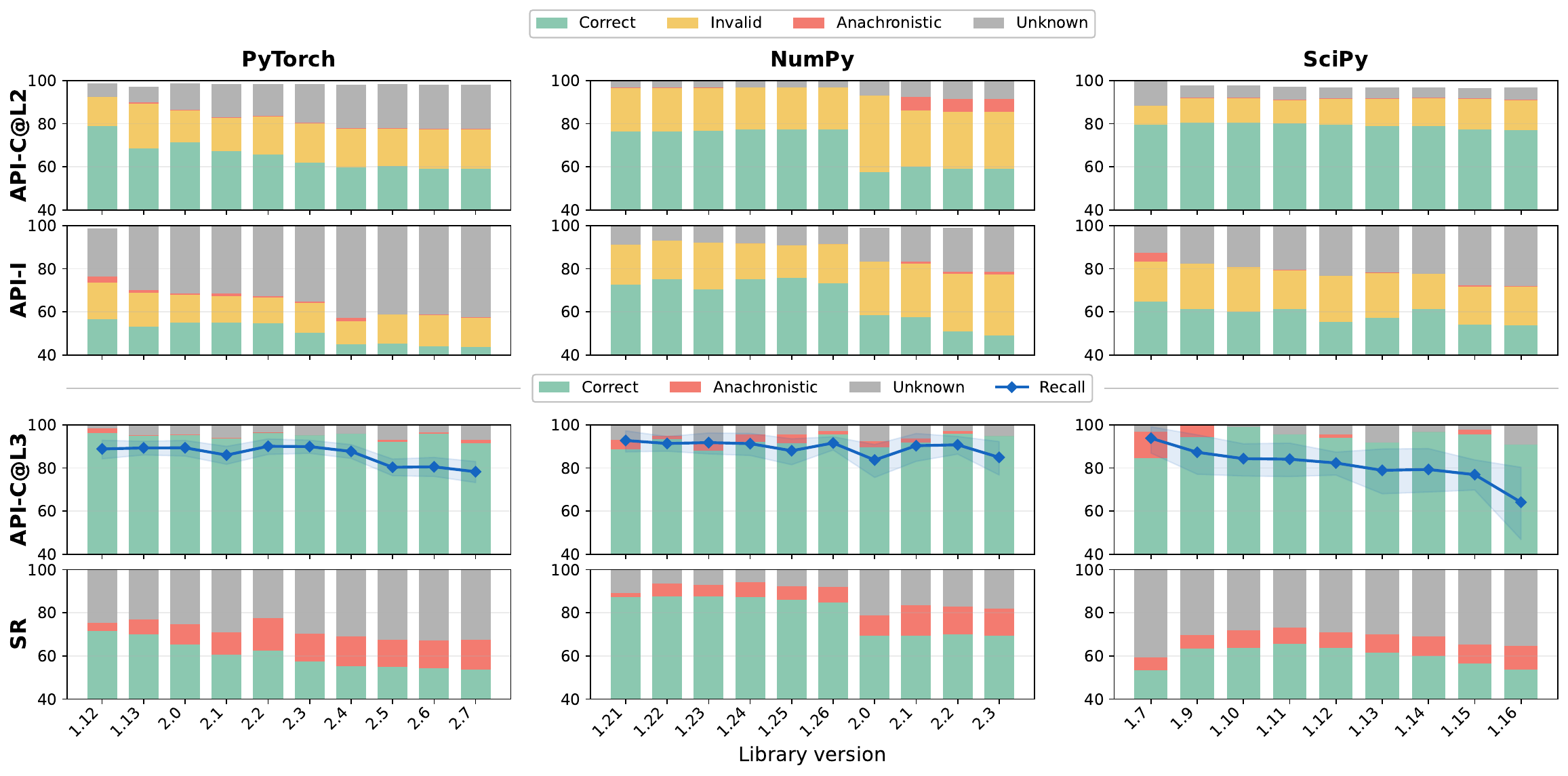}  
  \caption{Per-version taxonomy breakdown for \textbf{Gemini 2.5 Flash} across API-level (API Call with version \emph{API-C@L2}, API Identification \emph{API-I}) and parameter-level (API Call parameter recall \emph{API-C@L3}, Signature Recall \emph{SR}) predictions. The blue line reports parameter recall for \emph{API-C@L3}.}
  \label{fig:taxonomy-gemini-2.5}
\end{figure}

\begin{figure}[h]
  \centering
  \includegraphics[width=1\linewidth]{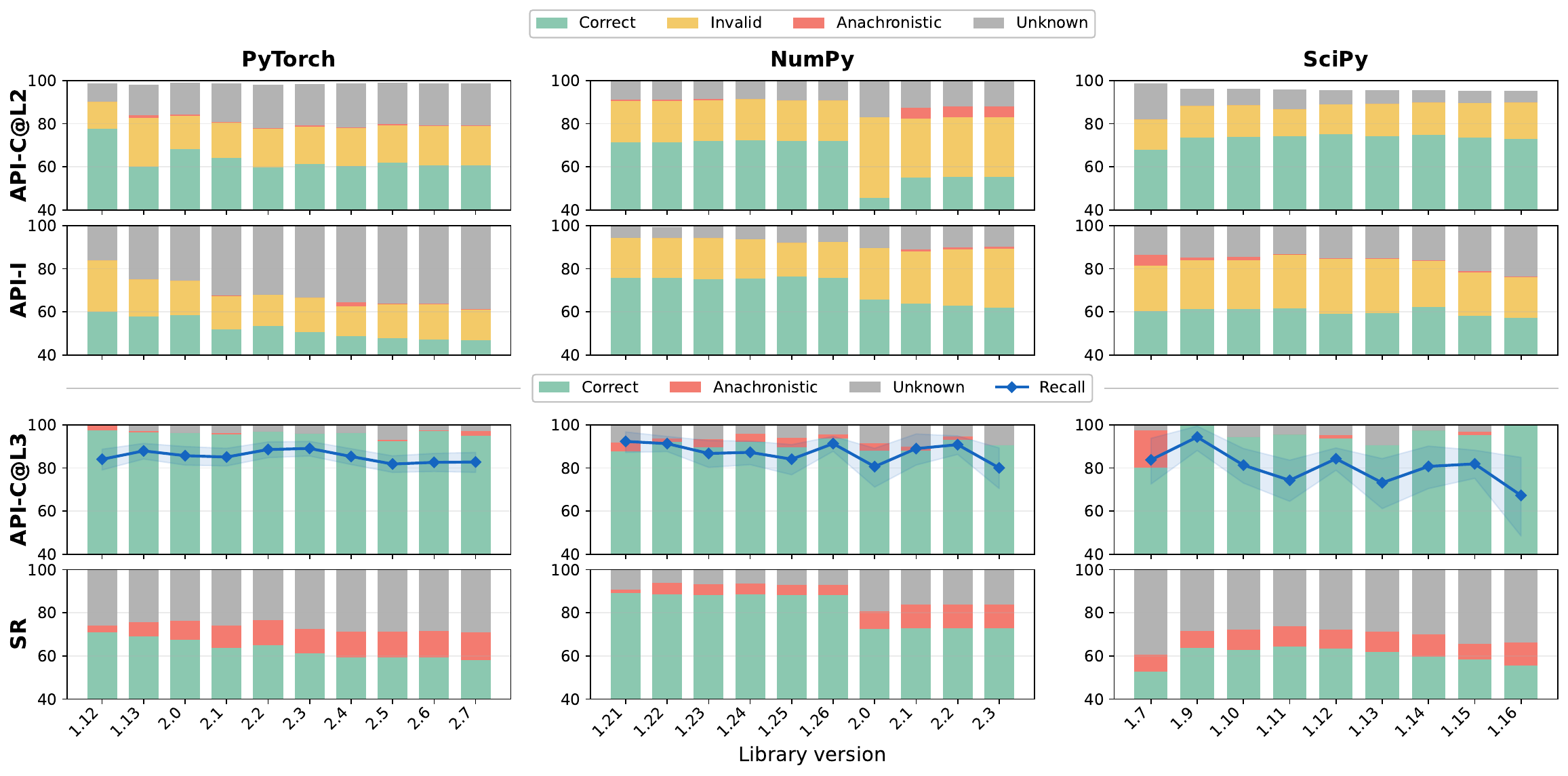}  

  \caption{Per-version taxonomy breakdown for \textbf{Qwen3.5 397B} across API-level (API Call with version \emph{API-C@L2}, API Identification \emph{API-I}) and parameter-level (API Call parameter recall \emph{API-C@L3}, Signature Recall \emph{SR}) predictions. The blue line reports parameter recall for \emph{API-C@L3}.}
  \label{fig:taxonomy-qwen-3.5-397b}
\end{figure}

This section provides the full taxonomy breakdown complementing the hallucination analysis in the main text. Table~\ref{tab:taxonomy} defines the complete category set at both API and parameter granularity. Figures~\ref{fig:taxonomy-sonnet-4}, \ref{fig:taxonomy-gemini-2.5}, and \ref{fig:taxonomy-qwen-3.5-397b} report the per-version stacked breakdown for Sonnet 4, Gemini 2.5 Flash, and Qwen3.5 397B respectively, extending the GPT-5.1 analysis shown in Figure~\ref{fig:taxonomy-gpt-5.1} of the main text. Tables~\ref{tab:taxonomy-apis} and \ref{tab:taxonomy-params} then report the aggregate taxonomy statistics at API and parameter granularity across all models and libraries.

\begin{table}[h]
\centering
\small
\caption{Hallucination taxonomy at API name and parameter granularity.}
\begin{tabular}{@{}c  l p{8.5cm}@{}}
\toprule
\textbf{Granularity} & \textbf{Category} & \textbf{Definition} \\
\midrule
\multirow{4}{*}{\rotatebox[origin=c]{90}{\textbf{API-level\qquad\qquad}}}
& Correct &
  The predicted API matches the ground-truth API at the target version. \\
\cmidrule(l){2-3}
& Invalid &
 An API that exists at the target version but does not match the intended one. Measures whether the model falls back to a plausible yet incorrect API when it fails to retrieve the correct one. \\
\cmidrule(l){2-3}
& Anachronistic &
  An API that exists in the library's history but is \emph{not} present at the target version: it may have been deprecated in a past release or introduced in a future one.
  \\
\cmidrule(l){2-3}
& Unknown &
  An API that does not appear in any of the observed versions of the library, most likely a non-existing API. \\

\midrule
\multirow{3}{*}{\rotatebox[origin=c]{90}{\textbf{Param-level\qquad}}}
& Correct &
  The parameter belongs to the ground-truth signature at the target version. \\
\cmidrule(l){2-3}
& Anachronistic &
  The parameter exists in API's signature history but not at the target version: it may have been removed in a past release or added in a future one. \\
\cmidrule(l){2-3}
& Unknown &
  The parameter has never appeared in any observed signature of this API, most likely a non-existing parameter. \\
\bottomrule
\end{tabular}
\label{tab:taxonomy}
\end{table}

\input{sections/tables/taxonomy_apis}
\input{sections/tables/taxonomy_params}

\clearpage
\subsection{Knowledge Cutoff Control}
\label{app:cutoff}
\begin{figure}[h]
\centering
\includegraphics[width=1\linewidth]{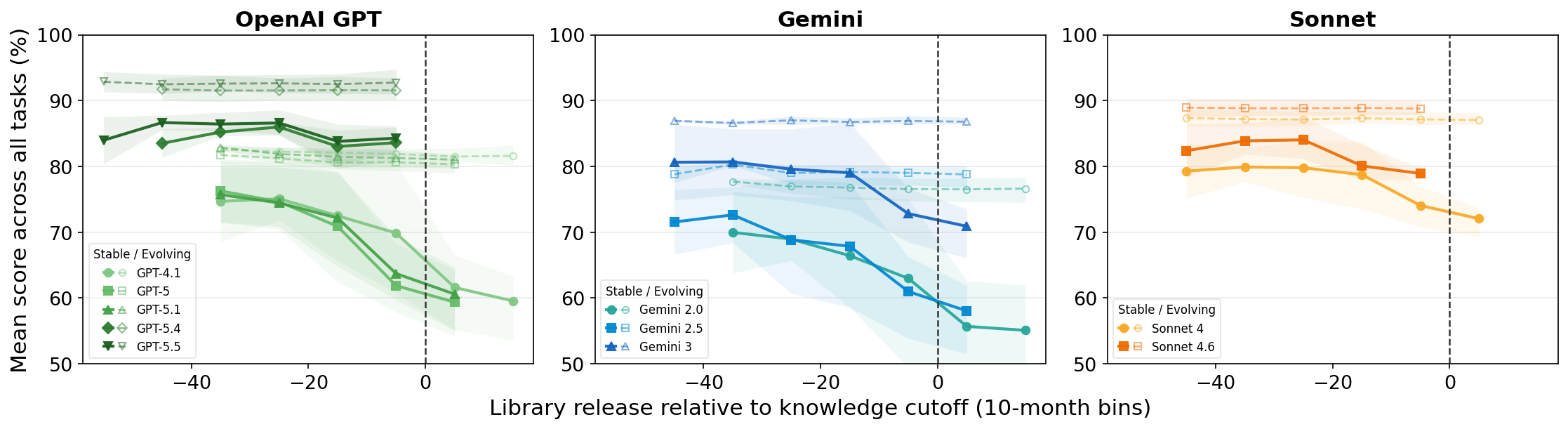}
\caption{Mean score across all tasks as a function of library release date relative to each model's knowledge cutoff (10-month bins; the dashed vertical line marks the cutoff). Stable APIs (dashed) remain flat, while evolving APIs (solid) decline, including for models whose cutoff postdates every benchmark version. Qwen3.5 is omitted because no cutoff is published.}
\label{fig:cutoff}
\end{figure}

A natural concern is that the degradation toward recent releases merely reflects proximity to each model's knowledge cutoff. To control for this, we collect the release dates of all 29 tracked library versions and the published knowledge cutoff of each evaluated model, and place every release relative to the cutoff of the model being evaluated. Releases are grouped into 10-month bins aligned so that the cutoff falls on a bin boundary, and stable and evolving APIs are reported separately (Figure~\ref{fig:cutoff}). The Qwen3.5 models are excluded: Alibaba does not publish a knowledge cutoff, and we omit them rather than estimate one.

GPT-5.4 and Sonnet~4.6 (cutoff August 2025) and GPT-5.5 (cutoff December 2025) postdate every benchmark version, yet they still show a stable/evolving gap that is visible at the oldest releases and grows toward the newest, while stable performance stays flat. For models whose cutoffs fall inside the version range, evolving performance drops by 4.8 points across bins that lie strictly before the cutoff, and by a further 8.3 points beyond it. Stable APIs act as a within-bin control: they come from the same releases, at the same distance from the cutoff, under the same prompts, and differ only in whether the API changed. If proximity to the cutoff drove the effect, both curves would fall; only the evolving curve does. We therefore attribute the decay not to whether a release falls before or after the cutoff, but to how much code using each version had accumulated by training time. Although Qwen3.5 cannot be placed on this axis, its three sizes share a training corpus, and therefore a cutoff, yet show almost identical decay curves (Figure~\ref{fig:families_row_all_norm}): capability moves the level of the curve, not its slope.

\subsection{Reasoning Budget Ablation}
\label{app:reasoning}
All main results are obtained without a thinking budget. To test this choice, we repeated the PyTorch evaluation, covering all tasks and augmentation levels, with reasoning enabled for GPT-5.1, Gemini~3 Flash, and Sonnet~4.6. The scope is limited by cost: a full run requires about 172k requests per model, and reasoning raises the cost by an order of magnitude.

\begin{figure}[h]
\centering
\includegraphics[width=1\linewidth]{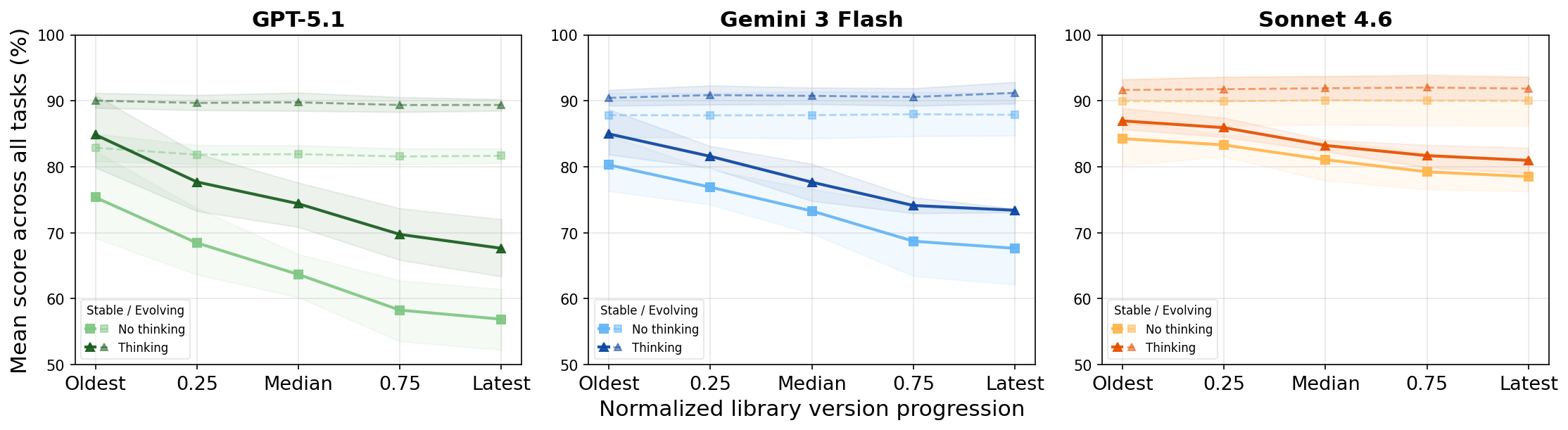}
\caption{Reasoning ablation on PyTorch: mean score across all tasks as a function of normalized library version progression, with and without reasoning, for GPT-5.1, Gemini~3 Flash, and Sonnet~4.6. Stable APIs are dashed and evolving APIs are solid.}
\label{fig:reasoning-mean}
\end{figure}

Reasoning improves absolute performance as expected, and it helps evolving APIs more than stable ones (Figure~\ref{fig:reasoning-mean}). It does not change the behavior described in the main text. The gap between stable and evolving APIs still widens at the same rate from the oldest to the newest release in both modes, and the shape of every curve is preserved. The reasoning budget lifts the curve without altering it.
For GPT-5.1, the gap grows from 7.5 to 24.8 points without reasoning and from 5.2 to 21.7 points with it; for Gemini~3 Flash, from 7.5 to 20.2 and from 5.5 to 17.8; for Sonnet~4.6, from 5.6 to 11.5 and from 4.7 to 10.9.

\begin{figure}[h]
\centering
\includegraphics[width=1\linewidth]{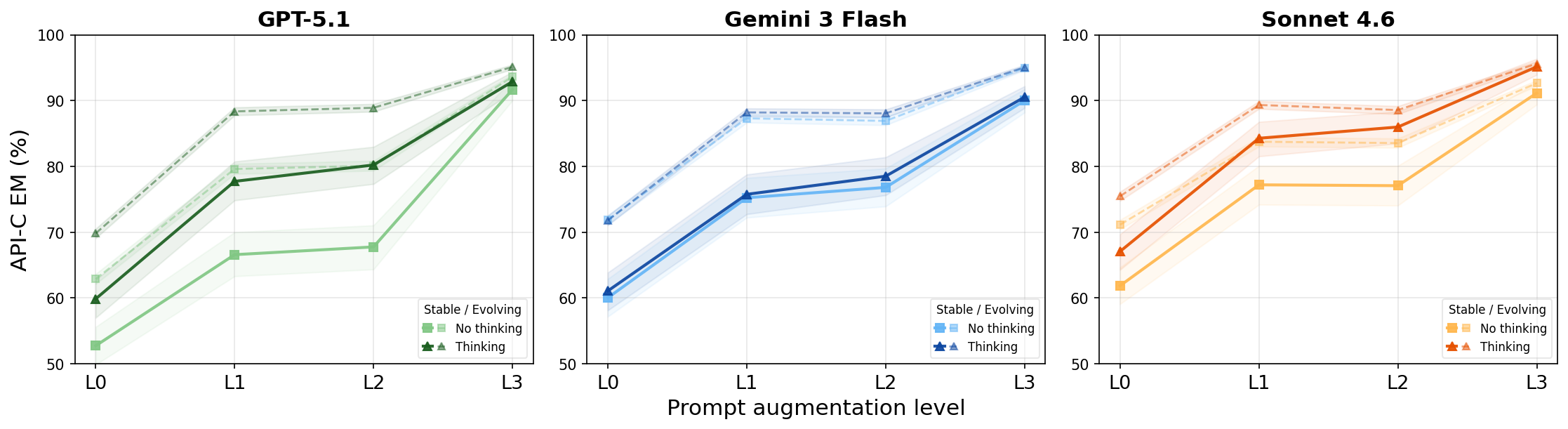}
\caption{Reasoning ablation on PyTorch: \emph{API-C} exact match per prompt augmentation level (L0 to L3), with and without reasoning. Adding the version constraint (L1 to L2) yields only marginal gains in both modes.}
\label{fig:reasoning-ladder}
\end{figure}

With a reasoning budget, the models do make some additional use of the version pin; the improvement is nonetheless marginal (Figure~\ref{fig:reasoning-ladder}).
Moving from L1 to L2 improves API-C accuracy on evolving APIs by 2.5, 2.8, and 1.7 points with reasoning, against 1.2, 1.6, and $-0.1$ points without.
API Identification and Signature Recall supply the target version in every prompt by construction, and they carry the largest gaps between stable and evolving APIs. Extended reasoning is therefore a compensation mechanism, not a solution: it buys accuracy without buying version awareness.

We also evaluated Sonnet~4.6 on API-C@L2 for PyTorch (16{,}444 instances) with a web-search tool available, using the same system prompt and thinking budget as above. The model did not invoke the tool once, despite every prompt containing the explicit version pin and despite having the reasoning budget to deliberate over it.

\clearpage
\subsection{Replication on Additional Libraries}
\label{app:replication}
\begin{figure}[h]
\centering
\includegraphics[width=1\linewidth]{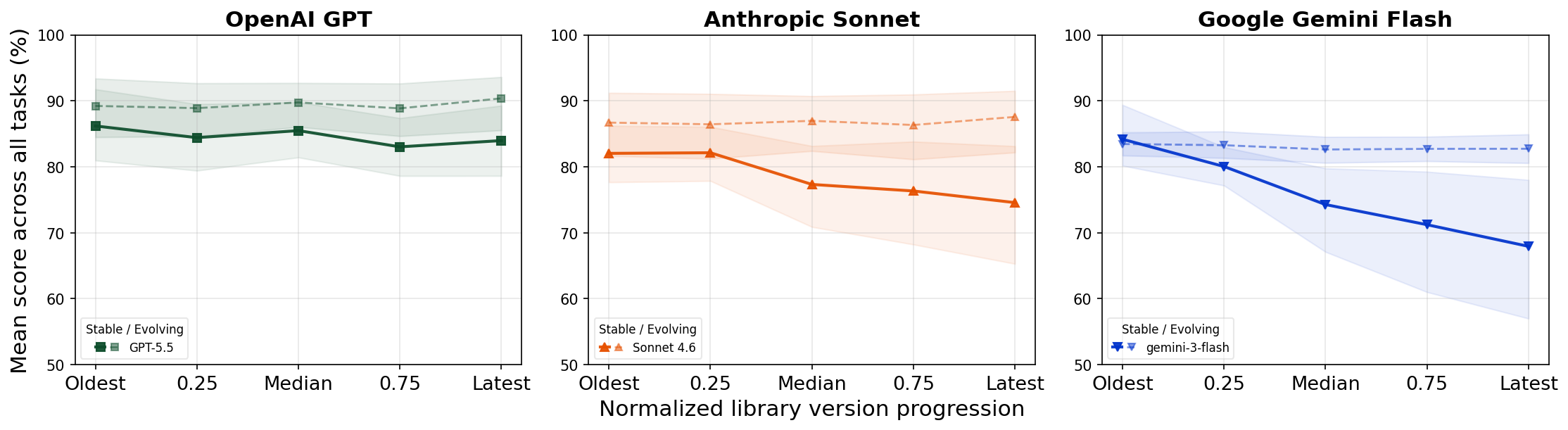}
\caption{Replication on Django, Transformers, and aiohttp (\emph{API-I} and \emph{SR}): mean score as a function of normalized library version progression for GPT-5.5, Sonnet~4.6, and Gemini~3 Flash. Stable APIs are dashed and evolving APIs are solid.}
\label{fig:replication}
\end{figure}

To assess external validity, we replicated the analysis on three widely used libraries outside our original criteria, Django, Transformers, and aiohttp, using API Identification and Signature Recall, which build directly from the compatibility matrices and do not require the code-mining pipeline. We evaluate GPT-5.5, Sonnet~4.6, and Gemini~3 Flash. These libraries offer less favorable conditions for the measurement than the original three: their covered histories are shorter and their evolving surfaces smaller, which is precisely why they did not meet our selection criteria. The effect appears in them regardless (Figure~\ref{fig:replication}). We leave the extension to other ecosystems and languages as future work.

%% file: sections/tables/results_appendix/results_with_ci.tex
\begin{sidewaystable}[h]
\centering
\caption{Per-task results stratified by API stability category. \emph{Stable} APIs remain
unchanged across all observed versions; \emph{Evolving} APIs vary within libraries versions.}
\label{tab:results-ci}
\footnotesize
\setlength{\tabcolsep}{2.5pt}
\resizebox{\columnwidth}{!}{%
\begin{tabular}{@{}lrrrrrrrrrrrrrrrrrr@{}}
\toprule
& \multicolumn{6}{c}{\textbf{PyTorch}} & \multicolumn{6}{c}{\textbf{NumPy}} & \multicolumn{6}{c}{\textbf{SciPy}} \\
\cmidrule(lr){2-7}\cmidrule(lr){8-13}\cmidrule(lr){14-19}
& \multicolumn{2}{c}{API-C@L1} & \multicolumn{2}{c}{API-I} & \multicolumn{2}{c}{SR} & \multicolumn{2}{c}{API-C@L1} & \multicolumn{2}{c}{API-I} & \multicolumn{2}{c}{SR} & \multicolumn{2}{c}{API-C@L1} & \multicolumn{2}{c}{API-I} & \multicolumn{2}{c}{SR} \\
\cmidrule(lr){2-3}\cmidrule(lr){4-5}\cmidrule(lr){6-7}\cmidrule(lr){8-9}\cmidrule(lr){10-11}\cmidrule(lr){12-13}\cmidrule(lr){14-15}\cmidrule(lr){16-17}\cmidrule(lr){18-19}
\textbf{Model} & \multicolumn{1}{c}{\tiny Stable} & \multicolumn{1}{c}{\tiny Evolving} & \multicolumn{1}{c}{\tiny Stable } & \multicolumn{1}{c}{\tiny Evolving} & \multicolumn{1}{c}{\tiny Stable } & \multicolumn{1}{c}{\tiny Evolving} & \multicolumn{1}{c}{\tiny Stable } & \multicolumn{1}{c}{\tiny Evolving} & \multicolumn{1}{c}{\tiny Stable } & \multicolumn{1}{c}{\tiny Evolving} & \multicolumn{1}{c}{\tiny Stable } & \multicolumn{1}{c}{\tiny Evolving} & \multicolumn{1}{c}{\tiny Stable } & \multicolumn{1}{c}{\tiny Evolving} & \multicolumn{1}{c}{\tiny Stable } & \multicolumn{1}{c}{\tiny Evolving} & \multicolumn{1}{c}{\tiny Stable } & \multicolumn{1}{c}{\tiny Evolving} \\
\midrule
GPT-4.1 & 75.8{\tiny$\pm$1.1} & 61.3{\tiny$\pm$3.3} & 85.3{\tiny$\pm$0.9} & 60.0{\tiny$\pm$1.2} & 86.8{\tiny$\pm$0.6} & 62.2{\tiny$\pm$0.6} & 75.8{\tiny$\pm$1.4} & 77.0{\tiny$\pm$3.2} & 80.9{\tiny$\pm$1.1} & 72.4{\tiny$\pm$2.2} & 84.8{\tiny$\pm$0.8} & 78.8{\tiny$\pm$0.9} & 88.9{\tiny$\pm$1.6} & 79.6{\tiny$\pm$4.2} & 81.3{\tiny$\pm$1.2} & 66.0{\tiny$\pm$1.3} & 85.0{\tiny$\pm$0.7} & 61.9{\tiny$\pm$0.7} \\
GPT-5 & 74.8{\tiny$\pm$1.1} & 66.4{\tiny$\pm$3.3} & 81.6{\tiny$\pm$1.0} & 52.7{\tiny$\pm$1.3} & 84.4{\tiny$\pm$0.6} & 62.8{\tiny$\pm$0.6} & 74.3{\tiny$\pm$1.4} & 72.8{\tiny$\pm$3.2} & 80.6{\tiny$\pm$1.1} & 72.6{\tiny$\pm$2.1} & 84.7{\tiny$\pm$0.8} & 79.7{\tiny$\pm$0.8} & 86.1{\tiny$\pm$1.7} & 78.3{\tiny$\pm$4.2} & 78.2{\tiny$\pm$1.3} & 65.2{\tiny$\pm$1.4} & 83.0{\tiny$\pm$0.8} & 61.2{\tiny$\pm$0.7} \\
GPT-5.1 & 76.3{\tiny$\pm$1.1} & 66.6{\tiny$\pm$3.4} & 83.6{\tiny$\pm$0.9} & 54.8{\tiny$\pm$1.2} & 83.9{\tiny$\pm$0.6} & 63.0{\tiny$\pm$0.6} & 74.4{\tiny$\pm$1.4} & 71.6{\tiny$\pm$3.5} & 81.9{\tiny$\pm$1.1} & 72.2{\tiny$\pm$2.2} & 85.3{\tiny$\pm$0.8} & 80.6{\tiny$\pm$0.8} & 88.1{\tiny$\pm$1.7} & 79.8{\tiny$\pm$4.1} & 78.1{\tiny$\pm$1.3} & 66.3{\tiny$\pm$1.4} & 84.7{\tiny$\pm$0.7} & 61.9{\tiny$\pm$0.7} \\
GPT-5.4 & 87.2{\tiny$\pm$0.8} & 85.3{\tiny$\pm$2.5} & 92.4{\tiny$\pm$0.6} & 86.1{\tiny$\pm$0.8} & 94.5{\tiny$\pm$0.4} & 85.8{\tiny$\pm$0.4} & 86.2{\tiny$\pm$1.2} & 80.2{\tiny$\pm$3.0} & 91.5{\tiny$\pm$0.8} & 84.9{\tiny$\pm$1.6} & 91.8{\tiny$\pm$0.7} & 87.7{\tiny$\pm$0.7} & 93.4{\tiny$\pm$1.3} & 89.0{\tiny$\pm$3.0} & 89.7{\tiny$\pm$0.9} & 86.5{\tiny$\pm$1.0} & 98.0{\tiny$\pm$0.3} & 81.2{\tiny$\pm$0.6} \\
GPT-5.5 & 90.4{\tiny$\pm$0.8} & 89.0{\tiny$\pm$2.2} & 92.3{\tiny$\pm$0.6} & 84.3{\tiny$\pm$0.9} & 95.7{\tiny$\pm$0.3} & 86.1{\tiny$\pm$0.4} & 88.0{\tiny$\pm$1.1} & 83.6{\tiny$\pm$2.9} & 92.2{\tiny$\pm$0.8} & 87.2{\tiny$\pm$1.6} & 92.5{\tiny$\pm$0.7} & 88.4{\tiny$\pm$0.7} & 95.4{\tiny$\pm$1.1} & 90.1{\tiny$\pm$3.0} & 89.3{\tiny$\pm$1.0} & 87.7{\tiny$\pm$0.9} & 99.0{\tiny$\pm$0.2} & 80.1{\tiny$\pm$0.6} \\
\addlinespace[2pt]
gemini-2.0-flash & 69.9{\tiny$\pm$1.1} & 54.3{\tiny$\pm$3.5} & 78.7{\tiny$\pm$1.0} & 38.6{\tiny$\pm$1.2} & 81.0{\tiny$\pm$0.6} & 58.1{\tiny$\pm$0.7} & 65.9{\tiny$\pm$1.6} & 72.3{\tiny$\pm$3.5} & 77.4{\tiny$\pm$1.2} & 64.7{\tiny$\pm$2.3} & 84.3{\tiny$\pm$0.8} & 78.5{\tiny$\pm$0.9} & 82.1{\tiny$\pm$1.9} & 75.9{\tiny$\pm$4.2} & 72.0{\tiny$\pm$1.4} & 58.7{\tiny$\pm$1.4} & 80.4{\tiny$\pm$0.8} & 56.9{\tiny$\pm$0.7} \\
gemini-2.5-flash & 76.6{\tiny$\pm$1.1} & 64.4{\tiny$\pm$3.3} & 83.0{\tiny$\pm$0.9} & 48.2{\tiny$\pm$1.2} & 85.0{\tiny$\pm$0.6} & 60.2{\tiny$\pm$0.6} & 71.9{\tiny$\pm$1.5} & 73.6{\tiny$\pm$3.4} & 76.2{\tiny$\pm$1.2} & 68.5{\tiny$\pm$2.2} & 85.1{\tiny$\pm$0.8} & 78.1{\tiny$\pm$0.9} & 84.4{\tiny$\pm$1.9} & 77.7{\tiny$\pm$4.3} & 72.3{\tiny$\pm$1.4} & 57.9{\tiny$\pm$1.4} & 80.6{\tiny$\pm$0.8} & 61.3{\tiny$\pm$0.7} \\
gemini-3-flash & 84.8{\tiny$\pm$0.9} & 75.2{\tiny$\pm$3.0} & 83.4{\tiny$\pm$0.9} & 63.7{\tiny$\pm$1.2} & 93.2{\tiny$\pm$0.4} & 76.1{\tiny$\pm$0.5} & 83.2{\tiny$\pm$1.3} & 81.3{\tiny$\pm$3.0} & 83.0{\tiny$\pm$1.1} & 78.5{\tiny$\pm$2.0} & 90.6{\tiny$\pm$0.7} & 87.6{\tiny$\pm$0.7} & 91.6{\tiny$\pm$1.4} & 86.4{\tiny$\pm$3.4} & 77.3{\tiny$\pm$1.3} & 67.8{\tiny$\pm$1.3} & 96.5{\tiny$\pm$0.4} & 73.8{\tiny$\pm$0.6} \\
\addlinespace[2pt]
Qwen3.5 35B & 50.6{\tiny$\pm$1.3} & 40.1{\tiny$\pm$3.5} & 59.8{\tiny$\pm$1.2} & 31.9{\tiny$\pm$1.1} & 75.3{\tiny$\pm$0.7} & 51.9{\tiny$\pm$0.7} & 47.6{\tiny$\pm$1.7} & 56.0{\tiny$\pm$3.8} & 48.2{\tiny$\pm$1.5} & 50.9{\tiny$\pm$2.4} & 67.0{\tiny$\pm$1.0} & 68.2{\tiny$\pm$1.0} & 59.9{\tiny$\pm$2.5} & 53.7{\tiny$\pm$5.1} & 49.8{\tiny$\pm$1.6} & 36.2{\tiny$\pm$1.3} & 56.1{\tiny$\pm$1.0} & 47.8{\tiny$\pm$0.7} \\
Qwen3.5 122B & 56.9{\tiny$\pm$1.2} & 48.4{\tiny$\pm$3.6} & 71.3{\tiny$\pm$1.1} & 46.1{\tiny$\pm$1.2} & 79.1{\tiny$\pm$0.7} & 57.9{\tiny$\pm$0.7} & 54.4{\tiny$\pm$1.7} & 53.2{\tiny$\pm$3.9} & 59.8{\tiny$\pm$1.4} & 63.8{\tiny$\pm$2.3} & 77.2{\tiny$\pm$1.0} & 75.8{\tiny$\pm$0.9} & 67.0{\tiny$\pm$2.3} & 65.4{\tiny$\pm$4.7} & 61.4{\tiny$\pm$1.6} & 50.8{\tiny$\pm$1.4} & 70.4{\tiny$\pm$1.0} & 56.2{\tiny$\pm$0.7} \\
Qwen3.5 397B & 70.4{\tiny$\pm$1.1} & 63.7{\tiny$\pm$3.3} & 80.2{\tiny$\pm$1.0} & 50.3{\tiny$\pm$1.3} & 85.7{\tiny$\pm$0.6} & 62.9{\tiny$\pm$0.6} & 67.8{\tiny$\pm$1.6} & 68.0{\tiny$\pm$3.6} & 77.2{\tiny$\pm$1.2} & 72.5{\tiny$\pm$2.2} & 84.8{\tiny$\pm$0.9} & 80.3{\tiny$\pm$0.9} & 80.1{\tiny$\pm$1.9} & 73.8{\tiny$\pm$4.7} & 75.1{\tiny$\pm$1.4} & 59.8{\tiny$\pm$1.4} & 83.2{\tiny$\pm$0.8} & 62.1{\tiny$\pm$0.7} \\
\addlinespace[2pt]
Sonnet 4 & 83.1{\tiny$\pm$1.0} & 74.2{\tiny$\pm$3.1} & 88.3{\tiny$\pm$0.8} & 73.5{\tiny$\pm$1.1} & 92.1{\tiny$\pm$0.5} & 75.1{\tiny$\pm$0.6} & 80.8{\tiny$\pm$1.3} & 77.1{\tiny$\pm$3.2} & 86.5{\tiny$\pm$0.9} & 79.4{\tiny$\pm$2.0} & 90.8{\tiny$\pm$0.7} & 86.2{\tiny$\pm$0.7} & 90.4{\tiny$\pm$1.5} & 84.3{\tiny$\pm$3.7} & 80.0{\tiny$\pm$1.2} & 75.3{\tiny$\pm$1.2} & 95.7{\tiny$\pm$0.4} & 73.8{\tiny$\pm$0.7} \\
Sonnet 4.6 & 81.9{\tiny$\pm$1.0} & 77.2{\tiny$\pm$2.9} & 90.7{\tiny$\pm$0.7} & 79.8{\tiny$\pm$1.0} & 94.8{\tiny$\pm$0.4} & 84.4{\tiny$\pm$0.4} & 79.8{\tiny$\pm$1.3} & 78.7{\tiny$\pm$3.1} & 90.0{\tiny$\pm$0.8} & 85.8{\tiny$\pm$1.7} & 92.3{\tiny$\pm$0.7} & 87.1{\tiny$\pm$0.7} & 84.9{\tiny$\pm$1.8} & 82.2{\tiny$\pm$3.8} & 84.4{\tiny$\pm$1.1} & 83.8{\tiny$\pm$1.0} & 97.9{\tiny$\pm$0.3} & 79.5{\tiny$\pm$0.6} \\
\bottomrule
\end{tabular}
}%
\end{sidewaystable}

%% file: sections/tables/taxonomy_apis.tex
\begin{table}
\centering
\caption{API-level taxonomy (API-C@L2 and API-I) on version-specific APIs. C=correct, I=invalid, A=anachronistic, U=unknown. 
Wrong-but-valid APIs dominate the error budget; anachronistic predictions are negligible.}
\label{tab:taxonomy-apis}
\footnotesize
\setlength{\tabcolsep}{2.2pt}
\resizebox{\columnwidth}{!}{%
\begin{tabular}{@{}lrrrrr rrrrr rrrrr rrrrr rrrrr rrrrr@{}}
\toprule
& \multicolumn{10}{c }{\textbf{PyTorch}} & \multicolumn{10}{c }{\textbf{NumPy}} & \multicolumn{10}{c}{\textbf{SciPy}} \\
\cmidrule(lr){2-11}\cmidrule(lr){12-21}\cmidrule(lr){22-31}
& \multicolumn{5}{c}{ API-C@L2} & \multicolumn{5}{c}{ API-I} & \multicolumn{5}{c}{ API-C@L2} & \multicolumn{5}{c }{ API-I} & \multicolumn{5}{c}{ API-C@L2} & \multicolumn{5}{c}{ API-I} \\
\cmidrule(lr){2-6}\cmidrule(lr){7-11}\cmidrule(lr){12-16}\cmidrule(lr){17-21}\cmidrule(lr){22-26}\cmidrule(lr){27-31}
\textbf{Model} & \multicolumn{1}{c}{\tiny Acc} & \multicolumn{1}{c}{\tiny C} & \multicolumn{1}{c}{\tiny I} & \multicolumn{1}{c}{\tiny A} & \multicolumn{1}{c}{\tiny U} & \multicolumn{1}{c}{\tiny Acc} & \multicolumn{1}{c}{\tiny C} & \multicolumn{1}{c}{\tiny I} & \multicolumn{1}{c}{\tiny A} & \multicolumn{1}{c}{\tiny U} & \multicolumn{1}{c}{\tiny Acc} & \multicolumn{1}{c}{\tiny C} & \multicolumn{1}{c}{\tiny I} & \multicolumn{1}{c}{\tiny A} & \multicolumn{1}{c}{\tiny U} & \multicolumn{1}{c}{\tiny Acc} & \multicolumn{1}{c}{\tiny C} & \multicolumn{1}{c}{\tiny I} & \multicolumn{1}{c}{\tiny A} & \multicolumn{1}{c }{\tiny U} & \multicolumn{1}{c}{\tiny Acc} & \multicolumn{1}{c}{\tiny C} & \multicolumn{1}{c}{\tiny I} & \multicolumn{1}{c}{\tiny A} & \multicolumn{1}{c}{\tiny U} & \multicolumn{1}{c}{\tiny Acc} & \multicolumn{1}{c}{\tiny C} & \multicolumn{1}{c}{\tiny I} & \multicolumn{1}{c}{\tiny A} & \multicolumn{1}{c}{\tiny U} \\
\midrule
GPT-4.1 & 62.9 & 62.9 & 20.1 & 0.1 & 15.6 & 60.0 & 60.0 & 10.1 & 0.6 & 29.3 & 77.1 & 77.1 & 15.8 & 0.9 & 6.2 & 72.4 & 72.4 & 19.1 & 0.4 & 8.1 & 80.6 & 80.6 & 10.2 & 0.0 & 7.3 & 66.0 & 66.0 & 15.2 & 0.2 & 18.5 \\
GPT-5 & 66.7 & 66.7 & 18.7 & 0.9 & 13.2 & 52.7 & 52.7 & 8.0 & 0.2 & 38.5 & 73.4 & 73.4 & 20.1 & 1.2 & 5.3 & 72.6 & 72.6 & 16.1 & 0.4 & 10.3 & 78.8 & 78.8 & 12.3 & 0.0 & 6.5 & 65.2 & 65.2 & 11.6 & 0.3 & 22.7 \\
GPT-5.1 & 67.8 & 67.8 & 15.1 & 0.5 & 15.9 & 54.8 & 54.8 & 10.0 & 0.4 & 34.2 & 72.3 & 72.3 & 21.3 & 1.1 & 5.3 & 72.2 & 72.2 & 18.2 & 0.5 & 9.1 & 77.7 & 77.7 & 13.6 & 0.3 & 6.0 & 66.3 & 66.3 & 12.8 & 0.3 & 20.3 \\
GPT-5.4 & 86.4 & 86.4 & 8.7 & 0.0 & 4.1 & 86.1 & 86.1 & 8.5 & 0.3 & 5.0 & 81.5 & 81.5 & 12.7 & 1.1 & 4.8 & 84.9 & 84.9 & 12.6 & 0.3 & 2.2 & 91.1 & 91.1 & 7.6 & 0.0 & 0.5 & 86.5 & 86.5 & 11.6 & 0.1 & 1.8 \\
GPT-5.5 & 90.4 & 90.4 & 5.1 & 0.4 & 3.7 & 84.3 & 84.3 & 8.0 & 0.5 & 6.7 & 84.5 & 84.5 & 13.0 & 1.1 & 1.2 & 87.2 & 87.2 & 11.1 & 0.2 & 1.5 & 89.8 & 89.8 & 8.4 & 0.0 & 0.5 & 87.7 & 87.7 & 8.2 & 0.1 & 3.8 \\
\addlinespace[2pt]
gemini-2.0-flash & 56.0 & 56.0 & 26.3 & 0.1 & 15.6 & 38.6 & 38.6 & 16.3 & 0.2 & 44.9 & 73.3 & 73.3 & 21.8 & 1.1 & 3.6 & 64.7 & 64.7 & 25.2 & 0.2 & 9.9 & 77.7 & 77.7 & 14.1 & 0.0 & 5.2 & 58.7 & 58.7 & 15.1 & 0.2 & 26.0 \\
gemini-2.5-flash & 61.3 & 61.3 & 17.8 & 0.3 & 18.9 & 48.2 & 48.2 & 13.4 & 0.7 & 37.5 & 74.2 & 74.2 & 21.0 & 0.9 & 3.7 & 68.5 & 68.5 & 20.2 & 0.2 & 10.9 & 77.7 & 77.7 & 12.3 & 0.3 & 7.1 & 57.9 & 57.9 & 19.2 & 0.2 & 22.6 \\
gemini-3-flash & 76.8 & 76.8 & 10.6 & 0.0 & 11.3 & 63.7 & 63.7 & 7.6 & 0.4 & 28.1 & 81.5 & 81.5 & 11.6 & 0.9 & 6.0 & 78.5 & 78.5 & 13.0 & 0.1 & 8.1 & 87.4 & 87.4 & 8.9 & 0.0 & 1.6 & 67.8 & 67.8 & 12.3 & 0.2 & 19.5 \\
\addlinespace[2pt]
Sonnet 4 & 76.3 & 76.3 & 15.3 & 0.0 & 7.7 & 73.5 & 73.5 & 10.9 & 0.3 & 14.9 & 76.8 & 76.8 & 17.5 & 1.2 & 4.5 & 79.4 & 79.4 & 14.9 & 0.4 & 5.3 & 84.3 & 84.3 & 9.4 & 0.0 & 3.4 & 75.3 & 75.3 & 14.7 & 0.2 & 9.6 \\
Sonnet 4.6 & 77.1 & 77.1 & 7.9 & 0.1 & 11.8 & 79.8 & 79.8 & 9.9 & 0.5 & 9.7 & 78.8 & 78.8 & 13.1 & 0.9 & 5.1 & 85.8 & 85.8 & 12.9 & 0.1 & 1.3 & 79.1 & 79.1 & 8.6 & 0.0 & 5.0 & 83.8 & 83.8 & 10.6 & 0.2 & 5.5 \\
\addlinespace[2pt]
Qwen3.5 35B & 40.0 & 40.0 & 25.8 & 2.1 & 30.3 & 31.9 & 31.9 & 14.4 & 0.8 & 52.9 & 57.0 & 57.0 & 26.9 & 2.5 & 13.0 & 50.9 & 50.9 & 29.2 & 0.2 & 19.7 & 52.1 & 52.1 & 28.3 & 0.0 & 16.0 & 36.2 & 36.2 & 20.6 & 0.3 & 42.8 \\
Qwen3.5 122B & 49.7 & 49.7 & 18.7 & 2.5 & 27.8 & 46.1 & 46.1 & 13.1 & 0.9 & 39.9 & 52.6 & 52.6 & 23.5 & 2.0 & 21.9 & 63.8 & 63.8 & 24.3 & 0.4 & 11.5 & 65.4 & 65.4 & 17.3 & 0.0 & 13.6 & 50.8 & 50.8 & 21.1 & 0.2 & 27.9 \\
Qwen3.5 397B & 62.6 & 62.6 & 17.4 & 0.4 & 18.3 & 50.3 & 50.3 & 15.6 & 0.3 & 33.8 & 69.4 & 69.4 & 20.2 & 1.1 & 9.3 & 72.5 & 72.5 & 19.8 & 0.2 & 7.4 & 72.0 & 72.0 & 15.7 & 0.0 & 8.4 & 59.8 & 59.8 & 22.6 & 0.6 & 17.0 \\
\bottomrule
\end{tabular}
}%
\vspace{1mm}
\end{table}

%% file: sections/tables/taxonomy_params.tex
\begin{table}
\centering
\caption{Parameter-level taxonomy (API-C@L3 and SR) on signature-changed APIs. 
  C=correct, A=anachronistic, U=unknown. Anachronistic parameters constitute a substantial fraction of SR errors.}
\label{tab:taxonomy-params}
\footnotesize
\setlength{\tabcolsep}{2.2pt}
\resizebox{\columnwidth}{!}{%
\begin{tabular}{@{}lrrrr rrrr rrrr rrrr rrrr rrrr@{}}
\toprule
& \multicolumn{8}{c }{\textbf{PyTorch}} & \multicolumn{8}{c }{\textbf{NumPy}} & \multicolumn{8}{c}{\textbf{SciPy}} \\
\cmidrule(lr){2-9}\cmidrule(lr){10-17}\cmidrule(lr){18-25}
& \multicolumn{4}{c}{API-C@L3} & \multicolumn{4}{c }{SR} & \multicolumn{4}{c}{API-C@L3} & \multicolumn{4}{c }{SR} & \multicolumn{4}{c}{API-C@L3} & \multicolumn{4}{c}{SR} \\
\cmidrule(lr){2-5}\cmidrule(lr){6-9}\cmidrule(lr){10-13}\cmidrule(lr){14-17}\cmidrule(lr){18-21}\cmidrule(lr){22-25}
\textbf{Model} & \multicolumn{1}{c}{\tiny Rec} & \multicolumn{1}{c}{\tiny C} & \multicolumn{1}{c}{\tiny A} & \multicolumn{1}{c}{\tiny U} & \multicolumn{1}{c}{\tiny F1} & \multicolumn{1}{c}{\tiny C} & \multicolumn{1}{c}{\tiny A} & \multicolumn{1}{c }{\tiny U} & \multicolumn{1}{c}{\tiny Rec} & \multicolumn{1}{c}{\tiny C} & \multicolumn{1}{c}{\tiny A} & \multicolumn{1}{c}{\tiny U} & \multicolumn{1}{c}{\tiny F1} & \multicolumn{1}{c}{\tiny C} & \multicolumn{1}{c}{\tiny A} & \multicolumn{1}{c }{\tiny U} & \multicolumn{1}{c}{\tiny Rec} & \multicolumn{1}{c}{\tiny C} & \multicolumn{1}{c}{\tiny A} & \multicolumn{1}{c}{\tiny U} & \multicolumn{1}{c}{\tiny F1} & \multicolumn{1}{c}{\tiny C} & \multicolumn{1}{c}{\tiny A} & \multicolumn{1}{c}{\tiny U} \\
\midrule
GPT-4.1 & 84.0 & 95.5 & 0.8 & 3.7 & 62.2 & 61.5 & 11.5 & 27.0 & 87.1 & 92.4 & 2.2 & 5.4 & 78.8 & 81.9 & 6.9 & 11.2 & 82.2 & 96.0 & 1.4 & 2.5 & 61.9 & 60.5 & 10.3 & 29.2 \\
GPT-5 & 84.1 & 95.3 & 1.0 & 3.7 & 62.8 & 61.8 & 11.7 & 26.5 & 89.1 & 91.5 & 2.3 & 6.2 & 79.7 & 82.4 & 7.1 & 10.4 & 84.0 & 94.6 & 1.6 & 3.7 & 61.2 & 59.4 & 10.9 & 29.7 \\
GPT-5.1 & 83.1 & 95.6 & 0.8 & 3.5 & 63.0 & 61.5 & 12.2 & 26.3 & 86.4 & 91.7 & 2.3 & 6.0 & 80.6 & 83.9 & 6.4 & 9.7 & 83.7 & 95.3 & 1.7 & 3.0 & 61.9 & 59.9 & 10.7 & 29.4 \\
GPT-5.4 & 90.1 & 95.8 & 1.4 & 2.8 & 85.8 & 86.8 & 6.4 & 6.8 & 92.4 & 93.0 & 2.3 & 4.8 & 87.7 & 88.7 & 5.1 & 6.2 & 87.9 & 95.9 & 2.5 & 1.6 & 81.2 & 81.1 & 3.7 & 15.2 \\
GPT-5.5 & 89.2 & 96.0 & 1.5 & 2.5 & 86.1 & 87.1 & 6.3 & 6.5 & 90.3 & 92.8 & 2.2 & 5.0 & 88.4 & 89.9 & 4.0 & 6.1 & 87.4 & 95.7 & 1.9 & 2.4 & 80.1 & 79.3 & 4.5 & 16.2 \\
\addlinespace[2pt]
gemini-2.0-flash & 85.1 & 93.8 & 1.3 & 5.0 & 58.1 & 58.2 & 10.1 & 31.7 & 89.4 & 92.4 & 2.4 & 5.2 & 78.5 & 81.6 & 6.6 & 11.7 & 83.1 & 94.8 & 1.5 & 3.7 & 56.9 & 55.4 & 11.3 & 33.3 \\
gemini-2.5-flash & 85.8 & 94.6 & 0.6 & 4.8 & 60.2 & 58.9 & 12.0 & 29.1 & 90.2 & 92.8 & 2.2 & 5.0 & 78.1 & 80.1 & 8.2 & 11.7 & 81.8 & 94.2 & 2.2 & 3.6 & 61.3 & 60.4 & 8.4 & 31.3 \\
gemini-3-flash & 86.6 & 95.9 & 0.7 & 3.5 & 76.1 & 76.3 & 9.5 & 14.1 & 89.6 & 92.8 & 2.3 & 4.8 & 87.6 & 88.7 & 5.4 & 6.0 & 87.5 & 95.4 & 2.5 & 2.1 & 73.8 & 72.9 & 6.4 & 20.7 \\
\addlinespace[2pt]
Sonnet 4 & 84.4 & 96.1 & 0.9 & 3.0 & 75.1 & 75.4 & 9.1 & 15.5 & 90.5 & 92.8 & 2.4 & 4.9 & 86.2 & 89.5 & 4.5 & 6.0 & 82.7 & 95.7 & 2.3 & 2.0 & 73.8 & 72.8 & 6.1 & 21.1 \\
Sonnet 4.6 & 89.6 & 96.7 & 0.6 & 2.7 & 84.4 & 85.3 & 6.8 & 7.9 & 92.2 & 93.1 & 2.2 & 4.8 & 87.1 & 89.4 & 4.6 & 6.0 & 86.3 & 94.7 & 2.5 & 2.9 & 79.5 & 78.3 & 5.6 & 16.0 \\
\addlinespace[2pt]
\addlinespace[2pt]
Qwen3.5 35B & 79.5 & 92.3 & 0.7 & 7.1 & 51.9 & 50.5 & 10.2 & 39.3 & 84.7 & 90.2 & 2.6 & 7.2 & 68.2 & 71.8 & 8.6 & 19.6 & 76.7 & 93.0 & 1.7 & 5.3 & 47.8 & 44.5 & 13.9 & 41.6 \\
Qwen3.5 122B & 82.2 & 95.0 & 0.6 & 4.4 & 57.9 & 57.1 & 9.8 & 33.1 & 86.4 & 91.0 & 2.1 & 6.9 & 75.8 & 78.6 & 7.3 & 14.1 & 79.8 & 94.4 & 1.7 & 3.9 & 56.2 & 55.1 & 9.8 & 35.0 \\
Qwen3.5 397B & 85.2 & 95.8 & 0.6 & 3.6 & 62.9 & 62.2 & 10.8 & 27.0 & 88.7 & 91.3 & 2.4 & 6.4 & 80.3 & 82.4 & 6.7 & 10.8 & 81.1 & 93.7 & 2.1 & 4.2 & 62.1 & 60.6 & 9.1 & 30.3 \\
\bottomrule
\end{tabular}
}%
\end{table}

%% file: sections/appendix/seus.tex
\section{Software Evolution Understanding Score}
\label{app:seus}

For each model, library, task, and version, let $S_v$ be the 
score on \emph{stable} APIs (unchanged across versions) and $E_v$ 
the score on \emph{evolving} APIs (version-specific or 
signature-changing). Both are averaged across the four task-level 
metrics (API-C@L2 exact match, API-C@L3 parameter recall, API-I 
exact match, and SR parameter F1) and combined via the harmonic 
mean $B_v = 2S_vE_v/(S_v+E_v)$. We also report evolution 
retention $E_v/S_v$ as a diagnostic.

The final model score averages across all 
$N = |\mathcal{L} \times \mathcal{T}|$ library--task pairs:
\[
\mathrm{SEUS} = \frac{1}{N} \sum_{(\ell,\, t)\, \in\, 
\mathcal{L} \times \mathcal{T}}
\left(
\mathbb{E}_v[B_v]
- \underbrace{\lambda \cdot \mathrm{Std}_v(E_v)}_{\text{stability penalty}}
- \underbrace{\gamma \cdot A_{\ell,t}}_{\text{anachronism penalty}}
\right)
\]
where $A_{\ell,t}$ is the anachronistic error rate on evolving 
APIs and $\lambda = \gamma = 0.5$. The stability penalty targets 
$E_v$ rather than $B_v$ because only variation in evolving APIs 
reflects genuine differences in version knowledge. The anachronism 
penalty captures a distinct failure mode: predictions that 
correspond to a real API from a different version, revealing 
active version confusion rather than mere lack of knowledge.

\begin{table}[h]
\centering
\caption{SEUS leaderboard. Stable and Evolving report average  task performance across the two API groups. 
Retention ($E_v/S_v$) measures how well a model preserves its performance as APIs 
evolve. Higher SEUS is better; the two penalties are subtracted 
from the balanced score.}
\label{tab:seus-leaderboard-full}
\footnotesize
\setlength{\tabcolsep}{2.2pt}
\resizebox{0.7\columnwidth}{!}{%
\begin{tabular}{@{}r l c c c  c c c }
\toprule
\textbf{Rank} & \textbf{Model} & \textbf{SEUS} & \multicolumn{3}{c}{\textbf{Performance}} & \multicolumn{2}{c}{\textbf{Penalty}}\\
\cmidrule(lr){4-6}\cmidrule(lr){7-8}
& & & \textbf{Stable} & \textbf{Evolving} & \textbf{Retention} & \textbf{Stability} & \textbf{Anach.} \\
\midrule
1 & GPT-5.4 & \textbf{86.0} & 92.0 & 86.4 & 94.1 & 2.0 & 1.0 \\
2 & GPT-5.5 & 85.1 & 92.3 & 86.1 & 93.5 & 2.7 & 0.9 \\
3 & Sonnet 4.6 & 81.0 & 89.3 & 82.4 & 92.4 & 3.4 & 1.0 \\
4 & Sonnet 4 & 77.6 & 87.8 & 78.5 & 89.6 & 3.7 & 1.1  \\
5 & gemini-3-flash & 77.3 & 87.8 & 78.5 & 89.4 & 3.9 & 1.2 \\
6 & GPT-5.1 & 70.7 & 83.8 & 71.6 & 85.3 & 4.2 & 1.5 \\
7 & GPT-4.1 & 70.6 & 84.3 & 71.7 & 85.0 & 4.5 & 1.5  \\
8 & GPT-5 & 70.0 & 83.0 & 71.5 & 86.0 & 4.3 & 1.6 \\
9 & gemini-2.5-flash & 68.9 & 81.6 & 69.8 & 85.4 & 4.1 & 1.5 \\
10 & Qwen3.5 397B & 67.9 & 80.2 & 68.9 & 85.9 & 3.9 & 1.4 \\
11 & gemini-2.0-flash & 66.0 & 79.7 & 67.5 & 84.4 & 4.5 & 1.5 \\
12 & Qwen3.5 122B & 60.3 & 70.5 & 62.3 & 87.8 & 3.5 & 1.6  \\
13 & Qwen3.5 35B & 52.6 & 63.3 & 54.7 & 85.9 & 3.4 & 1.8  \\
\bottomrule
\end{tabular}
}
\end{table}

Table~\ref{tab:seus-leaderboard-full} reports the full leaderboard. 
Several patterns stand out. First, the top two models 
(\textbf{GPT-5.4} and \textbf{GPT-5.5}) separate clearly from 
the rest, with SEUS scores above 85---driven by both high 
evolving-API performance and low penalties. Second, the gap 
between stable and evolving scores widens as overall performance 
decreases: top models retain over 93\% of their stable-API 
performance on evolving APIs, whereas weaker models drop below 
86\%. Third, the stability penalty is the primary differentiator 
among mid-tier models: for instance, \textbf{GPT-5.1} and 
\textbf{GPT-4.1} have nearly identical evolving scores (71.6 
vs.\ 71.7) but differ in version-to-version consistency (4.2 
vs.\ 4.5), which determines their final ordering. Finally, the 
anachronism penalty remains modest across all models (0.9--1.8\,pp), 
suggesting that version confusion is a secondary failure mode 
compared to outright performance degradation on evolving APIs.

\paragraph{SEUS Sensitivity to penalty calibration}
\label{app:seus-sensitivity}
We sweep $(\lambda, \gamma) \in \{0, 0.25, 0.5, 0.75, 1.0\}^2$ 
(25 settings) and find that Spearman rank correlation with the 
default setting never drops below 0.995, confirming that model 
ordering is driven by evolution-aware performance rather than 
penalty calibration.

The penalties do, however, differentiate individual models. For 
example, \textbf{GPT-4.1} ranks 7$^{th}$ at default but would 
overtake \textbf{GPT-5.1} 6$^{th}$ without the stability 
penalty. The two models have nearly identical balanced scores 
($\mathbb{E}_v[B_v]$: 76.6 vs.\ 76.5\,pp), but \textbf{GPT-4.1} 
exhibits higher version-to-version variance in evolving-API 
performance (9.0 vs.\ 8.4\,pp), which $\lambda$ correctly 
penalises.

%% file: sections/appendix/data_collection.tex
\section{Data Collection}
\label{app:data-collection}
The benchmark construction pipeline 
is \emph{fully automated}: it starts from a crawl of public GitHub repositories, joins their code with two independent views of each library's public API surface, and produces a unified, version-indexed dataset from which the three task-specific benchmarks are derived.
Every stage is deterministic and seeded; adding a new library version or a newly released library requires no manual 
curation, only re-running the pipeline against the updated release.

\subsection{Real Library Usage Collection.}
For each target library, we mine public GitHub repositories from a curated metadata index that links each repository to a specific pinned library version through its root-level dependency manifest (\texttt{requirements.txt}).
We keep only deduplicated repositories with a permissive license and at least ten stars, a threshold commonly used to filter low-quality or abandoned projects~\cite{bogomolov_long_2024}.
Each repository is cloned at the manifest commit and an AST walker scans every Python file, tracking imports, assignments, and alias propagation to identify all call sites that reach back to the target library.
Following the convention introduced by ~\citet{kuhar_libevolutioneval_2024}, we classify each call site as \emph{direct}, when the function expression is a pure attribute chain from the library root (e.g.\ \texttt{torch.nn.Linear()}), or \emph{indirect}, when access is mediated by an intermediate object or alias (e.g.\ \texttt{model.train()}).

\paragraph{Versioning and deduplication.}
Because each repository commits a dependency manifest pinning the exact library version, every mined call site inherits an unambiguous version label from its source repository, with no need for heuristic matching.
We apply aggressive deduplication to prevent any single snippet, API, or boilerplate pattern from dominating the dataset:
\textbf{(i)}~at the line level, when multiple call expressions share the same source line (e.g.\ \texttt{f(g(x))}), we retain only the outermost call, so that identical surface context is not counted twice;
\textbf{(ii)}~at the file level, only the first occurrence of each unqualified API name is kept, ensuring that repeated invocations of the same function inside a single file do not inflate its frequency.

\paragraph{Static API resolution.}
For each call site, we recover the fully qualified public name of the invoked API using the Jedi static analysis engine.
Jedi is run inside an isolated virtual environment created with \texttt{uv venv}, into which we install the \emph{exact} pinned library version of the source repository.
This ensures that attribute lookups, type inference, and module resolution all operate against the same API surface the developer saw at commit time, avoiding the version-drift issues that plague purely AST-based extractors.
The output of this stage is a raw cross-version usage database: every row links a code snippet, a pinned library version, and a resolved internal API path inside the versioned library.

\subsection{API Surface}
\paragraph{Public API inventory.}
We obtain the documented public surface of each library version from the official Sphinx \texttt{objects.inv} files, which library maintainers ship alongside their HTML documentation.
These inventories are discovered via PyPI metadata and versioned URL probing, and are parsed with \texttt{sphobjinv}.

\paragraph{Runtime introspection.}
In parallel with the inventory-based view, we install each library version into a dedicated environment and walk its module tree via \texttt{pkgutil} and runtime attribute inspection.
For every discovered entity, we record the \emph{canonical path} and the Python-internal identity 
that uniquely identifies the underlying object, regardless of how many aliases point to it.
Pairwise version comparison further classifies each API change as an addition, deletion, or parameter-level modification, producing a complete runtime record of the library's evolution.

\paragraph{Ground-truth construction.} Neither source alone is authoritative: documentation can list APIs that the introspection engine fails to resolve, and runtime introspection can surface private symbols that maintainers never meant to expose.
We therefore combine the two sources into a single ground truth.
An \emph{(API, version)} cell is marked \emph{Present} only when the API appears in \emph{both} the documentation inventory and the runtime output, ensuring that the symbol is both officially public and actually resolvable.
When only one source reports the API, the cell is marked \textbf{\textsc{na}} (ambiguous): this covers documented APIs that the introspection engine cannot extract, as well as runtime symbols absent from the official documentation.
A final \emph{deprecation pass} refines the boundary: if an API was \emph{Present} at some version but its documentation disappears in all subsequent versions while the runtime symbol persists, the trailing \textsc{na} cells are converted to \emph{Absent}, treating documentation removal as an intentional deprecation signal rather than an ambiguous gap.

\paragraph{Task-specific pool construction.}
The compatibility matrices drive the construction of all three task pools.
\emph{API Calling} (API-C) is built from the version-pinned code corpus; the raw pool is capped at 50 samples per \emph{(API, version)} cell to limit its size while maintaining high coverage across the version history.
The two declarative tasks are constructed directly from the matrices.
The \emph{API Identification} (API-I) raw pool enumerates every non-\textsc{na} cell of the API compatibility matrix~$\mathbf{A}$, yielding one identification query per \emph{(API, version)} pair with known existence status.
The \emph{Signature Recall} (SR) raw pool enumerates every active cell of the signature compatibility matrix~$\mathbf{S}$, retaining each \emph{(API, version, signature)} case for which the matrix records a valid signature.

\paragraph{Evaluation sampling.}
A deterministic sampler then balances each pool to keep the three tasks comparable while preserving coverage of version-sensitive behaviour.
Evolving APIs are retained in full across all tasks; a signature-lifetime strategy guarantees that every distinct lifetime of an API contributes to the final set.
Stable APIs form a controlled comparison group: a fixed subset is selected once (up to $C{=}1{,}000$ APIs per library, $m{=}5$ versions per API, $n{=}3$ snippets per cell for API-C) and propagated identically to API-C, API-I, and SR, ensuring the three tasks evaluate the same API population.
For API-C, samples are drawn uniformly across versions and code snippets per \emph{(API, version)} cell, preserving diversity across the version history rather than concentrating on versions with larger code corpora.
Further implementation details are available in the accompanying code release.

\begin{table}[h]
\centering
\caption{Data collection pipeline. \emph{Target} repos are those in our metadata snapshot whose \texttt{requirements.txt} pins the library; \emph{Cloned} counts repos successfully cloned and containing relevant Python files. \emph{\%Direct} is the fraction of direct (vs.\ indirect) API usages. \emph{API-C pool} is the pre-sampling benchmark; \emph{API-C eval} is the final stratified sample.}
\label{tab:pipeline}
\footnotesize
\setlength{\tabcolsep}{2.2pt}
\resizebox{\columnwidth}{!}{%
\begin{tabular}{@{}l  rr  rr r  rr  rr@{}}
\toprule
& \multicolumn{2}{c}{\textbf{Repositories}} & \multicolumn{3}{c}{\textbf{Raw usages}} & \multicolumn{2}{c}{\textbf{API-C pool}} & \multicolumn{2}{c@{}}{\textbf{API-C eval}} \\
\cmidrule(lr){2-3} \cmidrule(lr){4-6} \cmidrule(lr){7-8} \cmidrule(l){9-10}
\textbf{Library} & \textbf{Target} & \textbf{Cloned} & \textbf{Files} & \textbf{Usages} & \textbf{\%Direct} & \textbf{Samples} & \textbf{\%Direct} & \textbf{Samples} & \textbf{\%Direct} \\
\midrule
PyTorch & 4{,}259 & 3{,}715 & 134{,}277 & 1{,}435{,}824 & 41\% & 170{,}532 & 67\% & 16{,}444 & 63\% \\
NumPy & 4{,}765 & 3{,}592 & 73{,}073 & 302{,}013 & 82\% & 72{,}383 & 89\% & 9{,}275 & 82\% \\
SciPy & 3{,}914 & 2{,}037 & 10{,}184 & 23{,}761 & 19\% & 7{,}973 & 40\% & 3{,}948 & 42\% \\
\bottomrule
\end{tabular}
}
\end{table}

\begin{table}[h]
\centering
\caption{API surface statistics. \emph{APIs} is the total number of distinct public APIs tracked across all versions. \emph{Signatures} counts distinct (API, parameter-list) pairs; \emph{Multi-sig} is the number of APIs whose signature changed at least once. \emph{Doc} is the fraction of APIs with at least one scraped docstring.}
\label{tab:api-surface}
\begin{tabular}{@{}l r  rrrr  rr  r@{}}
\toprule
& & \multicolumn{4}{c}{\textbf{Entry types}} & \multicolumn{2}{c}{\textbf{Signatures}} & \\
\cmidrule(lr){3-6} \cmidrule(lr){7-8}
\textbf{Library} & \textbf{APIs} & \textbf{Func.} & \textbf{Meth.} & \textbf{Class} & \textbf{Prop.} & \textbf{Total} & \textbf{Multi} & \textbf{Doc} \\
\midrule
PyTorch & 5{,}178 & 1{,}519 & 2{,}080 & 1{,}245 & 334 & 7{,}171 & 1{,}184 & 96\% \\
NumPy & 2{,}566 & 1{,}054 & 995 & 181 & 336 & 2{,}998 & 407 & 95\% \\
SciPy & 3{,}642 & 1{,}288 & 1{,}631 & 370 & 353 & 4{,}921 & 792 & 95\% \\
\bottomrule
\end{tabular}
\end{table}

%% file: sections/appendix/doc_quality.tex
\section{Docstring Quality}
\label{app:doc-quality}

API-C prompt levels L1, L2 and API-I all rely on API docstrings.
Because not all docstrings are equally informative, we classify each API's documentation into a quality tier and restrict evaluation to tiers that carry genuine discriminative content.

\subsection{Source and Preprocessing}

For both API-C and API-I, docstrings are sourced from the API dictionary, which merges online Sphinx documentation with introspected docstrings for each (API, version) pair, preferring the online version when both are available.

Before classification and use in prompts, each raw docstring is processed through a pipeline of filters designed to retain only the descriptive core while removing any information that could trivially leak the answer.
Table~\ref{tab:doc-filters} lists each filter in application order.
\begin{table}[h]
\centering
\caption{Docstring preprocessing filters, applied in order.}
\small
\begin{tabular}{@{}l p{10cm}@{}}
\toprule
\textbf{Filter} & \textbf{Description} \\
\midrule
Signature stripping & Removes leading lines that reproduce the callable signature, common in NumPy-style docstrings. \\[2pt]
\midrule
Section truncation & Retains only the opening prose summary by truncating at the first structured section header. \\[2pt]
\midrule
Directive truncation & Truncates at the first reStructuredText directive, which typically follows the descriptive paragraph. \\[2pt]
\midrule
Length capping & Caps the result at 600 characters, breaking at the nearest word boundary. \\[2pt]
\midrule
Name redaction & Replaces all occurrences of the API's dotted path and its individual components with a placeholder token, preventing the model from reading the answer directly. \\[2pt]
\midrule
Collision deduplication & Removes APIs whose module prefix and redacted description are shared by multiple API names, since these prompts are no longer discriminative after name redaction. \\[2pt]
\midrule
Version scrubbing & Removes version-revealing phrases matched by seven regex patterns covering natural-language references, Sphinx version directives, and bare version literals. \\[2pt]
\midrule
Auto-generated placeholder & Descriptions produced by documentation tooling rather than written by a developer. \\[2pt]
\midrule
Non-discriminative & Descriptions under 50 characters that match boilerplate patterns shared across many APIs: getter/setter templates, forward references to other symbols, or vacuous meta-commentary. \\
\bottomrule
\end{tabular}
\label{tab:doc-filters}
\end{table}

To validate the efficacy of this filtering pipeline, we conducted a manual audit on a stratified sample of 500 retained short descriptions (under 50 characters). The audit confirmed that 98\% of the filtered descriptions successfully retained informative, discriminative semantics, with the remaining 2\% limited to acceptable noise such as ambiguous references.

\subsection{Quality Tiers}

\begin{table}[h]
\centering
\small
\caption{Accepted docstring quality tiers.}
\label{tab:doc-quality-defs}
\begin{tabular}{@{}l c p{9cm}@{}}
\toprule
\textbf{Tier} & \textbf{Length} & \textbf{Definition} \\
\midrule
\textbf{Short}  & $5<$chars$<50$     & A concise but content-bearing description that conveys the API's purpose. Short descriptions are included because, despite their brevity, they often suffice to distinguish an API from its neighbours (e.g.\ ``Apply a 2D transposed convolution''). \\[2pt]
\midrule
\textbf{Medium} & 50--99 chars   & A moderately detailed description, typically capturing the main behaviour and one or two qualifying details. \\[2pt]
\midrule
\textbf{Good}   & 100--299 chars & A detailed description that includes usage context, edge-case notes, or behavioural constraints. \\[2pt]
\midrule
\textbf{Rich}   & $\geq$300 chars & Comprehensive documentation with examples, mathematical formulations, or extensive parameter discussion. \\
\bottomrule
\end{tabular}
\end{table}

After applying the preprocessing pipeline and quality filters, the remaining instances are distributed across the four accepted tiers.
Tables~\ref{tab:doc-quality-t1} and~\ref{tab:doc-quality-t2} report this distribution for API-C and API-I respectively.
Both distributions reflect the final evaluation set after the sampling process.

\begin{table}
\centering
\small
\caption{Docstring quality distribution across API-C instances (L1 and L2 prompts).}
\label{tab:doc-quality-t1}
\begin{tabular}{@{}l r rrrr@{}}
\toprule
\textbf{Library} & \textbf{Total} & \textbf{Rich} & \textbf{Good} & \textbf{Medium} & \textbf{Short} \\
\midrule
PyTorch & 12{,}025 & 4{,}382 (36\%) & 4{,}077 (34\%) & 2{,}848 (24\%) & 718 (6\%) \\
NumPy & 7{,}955 & 1{,}918 (24\%) & 3{,}327 (42\%) & 2{,}102 (26\%) & 608 (8\%) \\
SciPy & 3{,}429 & 943 (28\%) & 1{,}348 (39\%) & 739 (22\%) & 399 (12\%) \\
\bottomrule
\end{tabular}

\caption{Docstring quality distribution across API-I instances.}
\label{tab:doc-quality-t2}
\begin{tabular}{@{}l r rrrr@{}}
\toprule
\textbf{Library} & \textbf{Total} & \textbf{Rich} & \textbf{Good} & \textbf{Medium} & \textbf{Short} \\
\midrule
PyTorch & 17{,}036 & 5{,}369 (32\%) & 6{,}716 (39\%) & 4{,}073 (24\%) & 878 (5\%) \\
NumPy & 9{,}385 & 2{,}259 (24\%) & 4{,}591 (49\%) & 1{,}997 (21\%) & 538 (6\%) \\
SciPy & 13{,}244 & 2{,}592 (20\%) & 6{,}130 (46\%) & 3{,}221 (24\%) & 1{,}301 (10\%) \\
\bottomrule
\end{tabular}
\end{table}

%% file: sections/appendix/prompts.tex
\section{Prompt Templates and Response Extraction}
\label{app:prompts}

This appendix lists the exact prompt templates used for each task and augmentation level, describes the response extraction pipeline, and reports extraction failure rates.
Placeholders in braces are filled at evaluation time from the benchmark data.

\subsection{API-C: API Calling}

\begin{promptbox}{API-C -- System message}
\consolasfont
You are a Python code completion assistant. Replace the <FILL\_HERE> marker with the correct code. Output ONLY the replacement text, nothing else.
\end{promptbox}

\begin{promptbox}{API-C@L0 -- Code only}
\consolasfont
\{prompt\}\{code\_str\}\{dot\}<FILL\_HERE>\\
\{suffix\}
\end{promptbox}

\begin{promptbox}{API-C@L1 -- Code + redacted description}
\consolasfont
\# Documentation\\
\# \{docstring\_redacted\}\\
\{prompt\}\{code\_str\}\{dot\}<FILL\_HERE>\\
\{suffix\}
\end{promptbox}

\begin{promptbox}{API-C@L2 -- Code + redacted description + version}
\consolasfont
\# requires: \{lib\}==\{ver\}\\
\# Documentation\\
\# \{docstring\_redacted\}\\
\{prompt\}\{code\_str\}\{dot\}<FILL\_HERE>\\
\{suffix\}
\end{promptbox}

\begin{promptbox}{API-C@L3 -- API name + version}
\consolasfont
\# requires: \{lib\}==\{ver\}\\
\# API: \{api\}\\
\{prompt\}\{code\_str\}\{dot\}<FILL\_HERE>\\
\{suffix\}
\end{promptbox}

\subsection{API-I: API Identification}

\begin{promptbox}{API-I -- System message}
\consolasfont
You are a Python API identification assistant. Given a module path prefix and a redacted description, identify the final API component that completes the path. Output ONLY the component name, nothing else.
\end{promptbox}

\begin{promptbox}{API-I -- User message}
\consolasfont
\{lib\} library, version \{version\}.\\
API path: \{prefix\}.\_\_\_\\
Description:\\
\{doc\}
\end{promptbox}

\subsection{SR: Signature Recall}

\begin{promptbox}{SR -- System message}
\consolasfont
You are a Python API signature expert. Given a fully qualified API name, library, and version, write its full function or method signature. Output ONLY the signature line (e.g. `func(a, b, c=None)`), nothing else.
\end{promptbox}

\begin{promptbox}{SR -- User message}
\consolasfont
Write the full function/method signature of `\{api\}` as it appears in \{lib\} \{version\}. Output ONLY the signature line, nothing else.
\end{promptbox}

\subsection{Response Extraction}
\label{app:extraction}

Each task applies a deterministic extraction pipeline to the raw model response before scoring.

\smartparagraph{API-C:}
The raw completion is passed through four post-processing steps:
(1)~markdown code fences are unwrapped if present,
(2)~inline backticks are stripped,
(3)~echo detection removes cases where the model repeats the input code prefix,
and (4)~verbose responses spanning more than two lines are truncated to the first expression.
The cleaned completion is then parsed to extract the API call: the fill portion is isolated, and the fully qualified API name is resolved by combining the call site with the code context.
A response is classified as a \emph{failure} if no API call can be extracted from the completion, which may occur when the model produces a different coding approach, prose commentary, or a syntactically incomplete expression.

\smartparagraph{API-I:}
The model is expected to return a single component name (e.g.\ \texttt{Dropout}).
The extraction pipeline strips backticks and whitespace, takes the first line if multiple are returned, discards any module prefix the model may have included, and validates that the result is a legal Python identifier.
The extracted leaf is then joined with the known module prefix to reconstruct the full API path.
A response is counted as a failure if no valid identifier can be extracted.

\smartparagraph{SR:}
The raw response is stripped of markdown fences and backticks, then normalized: the \texttt{def} keyword is removed if present, return type annotations are dropped, and parameter type annotations are stripped.
The parameter names are then extracted from the normalized signature using regex-based parsing.
A response is counted as a failure if the result is empty or if no parameter list can be parsed from it.

\subsection{Failure Rates}
Table~\ref{tab:fail-rates} reports the extraction failure rate for each task, model, and library. 
The Sonnet models shows slightly higher \textbf{API-C} failure rates due to thinking-token leakage - internal reasoning traces that bleed into the final response, breaking the expected output format. 
\textbf{API-I} failures are marginal and occur when the model produces an explanation instead of a single identifier. 
\textbf{SR} prompting and extraction are reliable.

\begin{table}[h]
\centering
\small
\caption{Response extraction failure rate (\%) per task, model, and library. For API-C, a failure indicates that the model produced a response but no API call could be extracted from it. For API-I and SR, a failure indicates an empty or unparseable response.}
\begin{tabular}{@{}ll @{\hspace{20pt}} rrrr @{\hspace{20pt}} r @{\hspace{20pt}} r@{}}
\toprule
& & \multicolumn{4}{c}{\textbf{API-C (by level)}} & \textbf{API-I} & \textbf{SR} \\
\cmidrule(lr){3-6}
\textbf{Model} & \textbf{Library} & \textbf{L0} & \textbf{L1} & \textbf{L2} & \textbf{L3} & & \\
\midrule
GPT-4.1 & PyTorch & 3.0\% & 1.3\% & 1.3\% & 0.9\% & $<$0.1\% & 0\% \\
GPT-4.1 & NumPy & 1.0\% & 0.5\% & 0.5\% & 0.4\% & 0\% & 0\% \\
GPT-4.1 & SciPy & 0.7\% & 0.5\% & 0.5\% & 0.3\% & 0\% & 0\% \\
GPT-5 & PyTorch & 1.7\% & 0.7\% & 0.7\% & 0.4\% & 0.3\% & 0\% \\
GPT-5 & NumPy & 1.0\% & 0.5\% & 0.6\% & 0.4\% & 0.3\% & 0\% \\
GPT-5 & SciPy & 0.4\% & 0.3\% & 0.4\% & 0.3\% & 0.5\% & 0\% \\
GPT-5.1 & PyTorch & 1.7\% & 0.6\% & 0.6\% & 0.4\% & 0.3\% & $<$0.1\% \\
GPT-5.1 & NumPy & 0.6\% & 0.5\% & 0.4\% & 0.3\% & 0.1\% & 0\% \\
GPT-5.1 & SciPy & 0.5\% & 0.2\% & 0.5\% & 0.2\% & 0.3\% & 0\% \\
GPT-5.4 & PyTorch & 1.5\% & 0.6\% & 0.4\% & 0.3\% & $<$0.1\% & 0\% \\
GPT-5.4 & NumPy & 0.9\% & 0.4\% & 0.3\% & 0.3\% & 0\% & 0\% \\
GPT-5.4 & SciPy & 0.5\% & 0.2\% & 0.2\% & 0.2\% & $<$0.1\% & 0\% \\
Claude Sonnet 4 & PyTorch & 1.6\% & 0.4\% & 0.4\% & 0.3\% & 0.2\% & 0\% \\
Claude Sonnet 4 & NumPy & 1.1\% & 0.4\% & 0.3\% & 0.3\% & 0.1\% & 0\% \\
Claude Sonnet 4 & SciPy & 0.7\% & 0.3\% & 0.4\% & 0.1\% & $<$0.1\% & 0\% \\
Claude Sonnet 4.6 & PyTorch & 4.1\% & 2.8\% & 2.8\% & 3.0\% & 0\% & 0\% \\
Claude Sonnet 4.6 & NumPy & 4.1\% & 3.8\% & 4.1\% & 3.7\% & 0\% & $<$0.1\% \\
Claude Sonnet 4.6 & SciPy & 3.5\% & 3.3\% & 4.3\% & 2.9\% & $<$0.1\% & $<$0.1\% \\
Gemini 2.0 Flash & PyTorch & 1.9\% & 1.1\% & 1.0\% & 0.8\% & $<$0.1\% & 0\% \\
Gemini 2.0 Flash & NumPy & 0.8\% & 0.7\% & 0.6\% & 0.4\% & $<$0.1\% & 0\% \\
Gemini 2.0 Flash & SciPy & 0.8\% & 0.4\% & 0.5\% & 0.3\% & 0.1\% & 0\% \\
Gemini 2.5 Flash & PyTorch & 1.8\% & 0.6\% & 0.6\% & 0.4\% & 0.1\% & 0\% \\
Gemini 2.5 Flash & NumPy & 0.7\% & 0.5\% & 0.5\% & 0.2\% & 0.1\% & $<$0.1\% \\
Gemini 2.5 Flash & SciPy & 0.6\% & 0.6\% & 0.5\% & 0.2\% & 0.1\% & 0\% \\
Gemini 3 Flash & PyTorch & 1.6\% & 0.6\% & 0.6\% & 0.5\% & 0.2\% & 0\% \\
Gemini 3 Flash & NumPy & 1.2\% & 0.6\% & 0.5\% & 0.5\% & 0.1\% & $<$0.1\% \\
Gemini 3 Flash & SciPy & 0.9\% & 0.3\% & 0.5\% & 0.3\% & 0.2\% & 0\% \\
Qwen3.5 35B & PyTorch & 2.5\% & 1.2\% & 1.2\% & 0.8\% & 0\% & 0\% \\
Qwen3.5 35B & NumPy & 3.4\% & 2.0\% & 2.2\% & 1.2\% & 0\% & 0\% \\
Qwen3.5 35B & SciPy & 1.5\% & 1.0\% & 1.1\% & 0.5\% & 0.1\% & 0\% \\
Qwen3.5 122B & PyTorch & 2.7\% & 1.3\% & 1.3\% & 0.8\% & 0\% & 0\% \\
Qwen3.5 122B & NumPy & 2.0\% & 1.1\% & 1.2\% & 0.7\% & 0\% & 0\% \\
Qwen3.5 122B & SciPy & 1.4\% & 1.0\% & 0.9\% & 0.7\% & 0\% & 0\% \\
Qwen3.5 397B & PyTorch & 2.3\% & 0.9\% & 1.0\% & 0.8\% & 0\% & 0\% \\
Qwen3.5 397B & NumPy & 1.3\% & 0.7\% & 0.7\% & 0.5\% & $<$0.1\% & 0\% \\
Qwen3.5 397B & SciPy & 0.8\% & 0.5\% & 0.6\% & 0.4\% & 0\% & 0\% \\
\bottomrule
\end{tabular}
\label{tab:fail-rates}
\end{table}

\subsection{Prompt Truncation Strategy}
\label{app:truncation}

Each code snippet is constructed from the source file prefix up to the cursor position, with a budget of 3{,}000 characters. 
Right context is capped at 3 to 10 lines with scope-boundary detection: the suffix terminates at the first \texttt{def}, \texttt{class}, or decorator at the same or lower indentation level.
If the full prefix exceeds the budget, the cascading strategy in Table~\ref{tab:truncation} is applied.

\begin{table}[h]
\centering
\small
\caption{Prompt truncation cascade. Steps are tried in order; the first to fit within the 3{,}000-character budget is used.}
\label{tab:truncation}
\begin{tabular}{@{}c l p{8cm}@{}}
\toprule
\textbf{\#} & \textbf{Strategy} & \textbf{Description} \\
\midrule
1 & Class scope & If the cursor is inside a class, the prefix is cut to start at the \texttt{class} line with a note \texttt{``(previous context omitted)''}. \\[2pt]
\midrule
2 & Method removal & Other methods in the class are removed one by one, starting from the farthest from the target. Method ranges are decorator-aware: each span includes leading decorators and trailing blank lines. This continues until the budget is met or all methods are removed. \\[2pt]
\midrule
3 & Function scope & If the cursor is inside a standalone function (not a class), the prefix is cut to start at the \texttt{def} line. \\[2pt]
\midrule
4 & Discard & If no strategy achieves the budget, the sample is discarded. \\
\bottomrule
\end{tabular}
\end{table}

When truncation occurs, library-relevant imports are re-attached: up to 8 import lines are prepended, filtered to only those whose imported names actually appear in the code context. Final dimension distribution in table~\ref{tab:snippet-lengths}.

\begin{table}[h]
\centering
\small
\caption{Distribution of API-C code snippet lengths. Budget-capped at 3{,}000 characters. Suffix is the right context after the cursor. Median and interquartile range (IQR, Q25 - Q75) are reported.}
\begin{tabular}{@{}l r | rr rr | rr rr @{}}
\toprule
& & \multicolumn{4}{c|}{\textbf{Prefix}} & \multicolumn{4}{c}{\textbf{Suffix}} \\
\cmidrule(lr){3-6} \cmidrule(lr){7-10}
\textbf{Library} & \textbf{$n$} & \textbf{Chars} & \textbf{IQR} & \textbf{Lines} & \textbf{IQR} & \textbf{Chars} & \textbf{IQR} & \textbf{Lines} & \textbf{IQR} \\
\midrule
PyTorch & \num{16444} & \num{1481} & \num{744}--\num{2303} & 44 & 24--63 & \num{301} & \num{147}--\num{414} & 9 & 4--10 \\
NumPy & \num{9275} & \num{1436} & \num{664}--\num{2333} & 44 & 23--67 & \num{284} & \num{150}--\num{391} & 9 & 4--10 \\
SciPy & \num{3948} & \num{1323} & \num{555}--\num{2274} & 41 & 19--64 & \num{288} & \num{150}--\num{398} & 9 & 4--10 \\
\bottomrule
\end{tabular}

\label{tab:snippet-lengths}
\end{table}

%% file: sections/appendix/examples.tex
\newcommand{\hlbox}[1]{\colorbox{yellow!40}{\texttt{#1}}}

\section{Prompt Examples}
\label{app:prompt-examples}

The following boxes show real prompts from the benchmark, one per library for each task.
Each prompt is the exact message sent to the model at evaluation time.

\subsubsection*{API-C: API Calling (L0)}
\label{app:prompt-examples-t1}

\begin{promptbox}{API-C@L0 Example -- PyTorch (target: \texttt{torch.compile}, v2.2)}
\consolasfont
\footnotesize
from contextlib import nullcontext\\
import copy\\
from pathlib import Path\\
import time\\
import yaml\\
\mbox{}\\
import torch\\
import wandb\\
\mbox{}\\
from logger.logger import DynamicsLogger\\
from optim.weight\_averaging import (\\
\hspace*{2em}WeightAverager,\\
\hspace*{2em}eval\_ema,\\
\hspace*{2em}eval\_wa,\\
\hspace*{2em}ExponentialWeightAverager,\\
)\\
from .utils import (\\
\hspace*{2em}eval,\\
\hspace*{2em}get\_batch,\\
\hspace*{2em}load\_checkpoint,\\
\hspace*{2em}load\_worker\_state,\\
\hspace*{2em}save\_checkpoint,\\
\hspace*{2em}save\_worker\_state,\\
)\\
\mbox{}\\
\mbox{}\\
def train(\\
\hspace*{2em}model,\\
\hspace*{2em}opt,\\
\hspace*{2em}datareaders,\\
\hspace*{2em}scheduler,\\
\hspace*{2em}exp\_dir,\\
\hspace*{2em}distributed\_backend,\\
\hspace*{2em}cfg,\\
):\\
\hspace*{2em}not\_compiled\_model = model\\
\hspace*{2em}if cfg.compile:\\
\hspace*{4em}print(f"Compiling model ...")\\
\hspace*{4em}model = torch.\hlbox{<FILL\_HERE>}\\
\hspace*{2em}if "cuda" in cfg.device:\\
\hspace*{4em}type\_ctx = torch.amp.autocast(\\
\hspace*{6em}device\_type="cuda",\\
\hspace*{6em}dtype=\{
\end{promptbox}

\subsubsection*{API-I: API Identification}

\begin{promptbox}{API-I Example -- SciPy (target: \texttt{ifht}, v1.7)}
\consolasfont
\footnotesize
scipy library, version 1.7.\\
API path: scipy.fft.\_\_\_\\
Description:\\
Compute the inverse fast Hankel transform.\\
\mbox{}\\
Computes the discrete inverse Hankel transform of a logarithmically spaced periodic sequence. This is the inverse operation to fht.
\end{promptbox}

\subsubsection*{SR: Signature Recall }

\begin{promptbox}{SR Example -- PyTorch v2.3 \\GT: \texttt{LinearReLU(in\_features, out\_features, bias=True, qconfig=None)}}
\consolasfont
\footnotesize
Write the full function/method signature of `torch.ao.nn.intrinsic.qat.LinearReLU` as it appears in torch 2.3. Output ONLY the signature line, nothing else.
\end{promptbox}

\section{Reproducibility}
\label{app:reproducibility}

The full benchmark dataset is publicly available at \url{https://huggingface.co/datasets/Anonymous-999/LibEvoBench}.
The evaluation code, pipeline implementation, and scripts to reproduce all results reported
in this paper are released at \url{https://zenodo.org/records/20066456}.
The dataset is also available at \url{https://huggingface.co/datasets/JetBrains/LibEvoBench},
and the code at \url{https://github.com/JetBrains-Research/LibEvoBench}.

%% file: checklist.tex
\section*{NeurIPS Paper Checklist}

\begin{enumerate}

\item {\bf Claims}
    \item[] Question: Do the main claims made in the abstract and introduction accurately reflect the paper's contributions and scope?
    \item[] Answer: \answerYes{} %
    \item[] Justification: The abstract and introduction clearly define the LibEvoBench dataset, the SEUS metric, and the core findings regarding models' version-obliviousness, which are well-supported by the experiments detailed in Sections \ref{sec:benchmark} and \ref{sec:results}.

\item {\bf Limitations}
    \item[] Question: Does the paper discuss the limitations of the work performed by the authors?
    \item[] Answer: \answerYes{} %
    \item[] Justification: The authors dedicate Section \ref{sec:limitations} to discuss limitations, specifically the zero thinking budget constraint and the potential data overlap between the mined open-source repositories and the models' training corpora.

\item {\bf Theory assumptions and proofs}
    \item[] Question: For each theoretical result, does the paper provide the full set of assumptions and a complete (and correct) proof?
    \item[] Answer: \answerNA{} %
    \item[] Justification: The paper focuses on empirical evaluation, introducing a new benchmark and metric; it does not introduce mathematical theorems or theoretical proofs.

    \item {\bf Experimental result reproducibility}
    \item[] Question: Does the paper fully disclose all the information needed to reproduce the main experimental results of the paper to the extent that it affects the main claims and/or conclusions of the paper (regardless of whether the code and data are provided or not)?
    \item[] Answer: \answerYes{} %
    \item[] Justification: The paper thoroughly explains the data collection methodology, sampling pipelines, models tested, prompt templates (Appendix \ref{app:prompts}), and evaluation parameters (e.g., zero temperature).

\item {\bf Open access to data and code}
    \item[] Question: Does the paper provide open access to the data and code, with sufficient instructions to faithfully reproduce the main experimental results, as described in supplemental material?
    \item[] Answer: \answerYes{} %
    \item[] Justification: The complete source code and associated datasets required to fully reproduce both the data collection methodology and the evaluation pipelines are publicly available: the dataset on Hugging Face and the code on GitHub and Zenodo (Appendix~\ref{app:reproducibility}).

\item {\bf Experimental setting/details}
    \item[] Question: Does the paper specify all the training and test details (e.g., data splits, hyperparameters, how they were chosen, type of optimizer) necessary to understand the results?
    \item[] Answer: \answerYes{} %
    \item[] Justification: The paper explicitly details all experimental settings required to understand and reproduce the results. This includes specific model inference parameters and exact prompt templates. 

\item {\bf Experiment statistical significance}
    \item[] Question: Does the paper report error bars suitably and correctly defined or other appropriate information about the statistical significance of the experiments?
    \item[] Answer: \answerYes{} %
    \item[] Justification: The paper includes shaded error bounds representing statistical spread in the main body plots and reports 95\% bootstrap confidence intervals in Table \ref{tab:results-ci}.

\item {\bf Experiments compute resources}
    \item[] Question: For each experiment, does the paper provide sufficient information on the computer resources (type of compute workers, memory, time of execution) needed to reproduce the experiments?
    \item[] Answer: \answerYes{} %
    \item[] Justification: The paper details the compute infrastructure in the Appendix \ref{app:infra}.
    
\item {\bf Code of ethics}
    \item[] Question: Does the research conducted in the paper conform, in every respect, with the NeurIPS Code of Ethics \url{https://neurips.cc/public/EthicsGuidelines}?
    \item[] Answer: \answerYes{} %
    \item[] Justification: We confirm that our research fully complies with the NeurIPS Code of Ethics and our data collection is limited to public, permissively licensed repositories.

\item {\bf Broader impacts}
    \item[] Question: Does the paper discuss both potential positive societal impacts and negative societal impacts of the work performed?
    \item[] Answer: \answerNA{} %
    \item[] Justification: As this research focuses strictly on benchmark methodology, it does not present any direct societal consequences.
    
\item {\bf Safeguards}
    \item[] Question: Does the paper describe safeguards that have been put in place for responsible release of data or models that have a high risk for misuse (e.g., pre-trained language models, image generators, or scraped datasets)?
    \item[] Answer: \answerNA{} %
    \item[] Justification: The released benchmark dataset strictly evaluates the technical capabilities of models using public software, posing no risk of malicious misuse.

\item {\bf Licenses for existing assets}
    \item[] Question: Are the creators or original owners of assets (e.g., code, data, models), used in the paper, properly credited and are the license and terms of use explicitly mentioned and properly respected?
    \item[] Answer: \answerYes{} %
    \item[] Justification: We strictly limit our data collection to repositories with permissive licenses. Because each source repository has its own specific license, we track and propagate this licensing information alongside the extracted data to ensure all terms of use are respected. All evaluated models and libraries are also properly cited.

\item {\bf New assets}
    \item[] Question: Are new assets introduced in the paper well documented and is the documentation provided alongside the assets?
    \item[] Answer: \answerYes{} %
    \item[] Justification: The newly introduced benchmark (\namebenchmark{}) and its accompanying codebase are thoroughly documented. We provide comprehensive usage instructions, data schemas, and reproduction scripts directly alongside the released data and code assets.

\item {\bf Crowdsourcing and research with human subjects}
    \item[] Question: For crowdsourcing experiments and research with human subjects, does the paper include the full text of instructions given to participants and screenshots, if applicable, as well as details about compensation (if any)? 
    \item[] Answer: \answerNA{} %
    \item[] Justification: We do not conduct crowdsourcing or research involving human subjects. Our empirical evaluation relies on automated code analysis and machine learning models.

\item {\bf Institutional review board (IRB) approvals or equivalent for research with human subjects}
    \item[] Question: Does the paper describe potential risks incurred by study participants, whether such risks were disclosed to the subjects, and whether Institutional Review Board (IRB) approvals (or an equivalent approval/review based on the requirements of your country or institution) were obtained?
    \item[] Answer: \answerNA{} %
    \item[] Justification: No human subjects were involved in the research.

\item {\bf Declaration of LLM usage}
    \item[] Question: Does the paper describe the usage of LLMs if it is an important, original, or non-standard component of the core methods in this research? Note that if the LLM is used only for writing, editing, or formatting purposes and does \emph{not} impact the core methodology, scientific rigor, or originality of the research, declaration is not required.
    \item[] Answer: \answerNA{} %
    \item[] Justification: We did not use LLMs as a component of our core method development or dataset generation. The language models discussed in the paper are the subjects being evaluated, not components of the benchmark construction methodology.

\end{enumerate}